# Multiscale Dynamics of Heatwave Persistence and Intensity Under Climate Change

Md Tasim Ferdous[1], Reda Snaiki[1], Abdelatif Merabtine[1 *]

[1] *Department of Construction Engineering, École de Technologie Supérieure, Université du Québec, Montréal, Québec, Canada*

**Corresponding author: abdelatif.merabtine@etsmtl.ca*

**Abstract:** Climate change is expected to intensify heatwave risk, yet increases in exceedance frequency alone do not explain why some regions experience disproportionately stronger amplification in event persistence. This study develops an integrated event–dynamical workflow to diagnose how warm-season heatwave characteristics evolve and to attribute projected changes to coherent, multiscale structures of temperature variability. Heatwaves are identified over southern Canada using a fixed historical percentile threshold (90th; reference 2001–2010, 15-day moving window) and a minimum-duration constraint and are summarized by frequency (HWF, HWN), persistence (HWMD, HWD), and intensity (HWI, HWM). The dynamical organization of the daily mean temperature field is characterized using multiresolution dynamic mode decomposition (mrDMD). Event and dynamical perspectives are linked through (i) heatwave-conditioned mode participation ratios that identify modes preferentially activated during heatwave days and (ii) spatial alignment analyses comparing mode-energy footprints to gridded heatwave-metric maps using hotspot overlap and Spearman rank association. Applied to CORDEX-NAM12 regional simulations (CRCM5 downscaling of CanESM5) under SSP5–8.5 for 2016–2025, 2051–2060, and 2091–2100, the results indicate a pronounced shift toward persistence-dominated heatwave regimes in the continental interior. Late-century increases in seasonal heatwave days are accompanied by substantially longer events, with regional HWMD reaching up to ~26.66 days/event and HWD up to ~69 days, alongside marked intensification of above-threshold heatwave intensity, with regional HWI reaching up to ~6.88 K). Dynamical diagnostics show a redistribution of dominant activity toward lower-frequency levels and weaker effective damping in interior regions, whereas coastal and maritime regions exhibit smaller shifts. Heatwave-relevant low-frequency modes display sustained activation during long events and spatial footprints that align with persistence and intensity hotspots, supporting a process-informed interpretation of regional heatwave amplification under climate change.

## 1. Introduction

Heatwaves are among the most consequential climate hazards, with impacts that extend from acute heat-related mortality and morbidity to cascading stresses on energy systems, transportation, water resources, and ecosystem function (Ebi et al., 2021; Kirchmeier-Young et al., 2025; Smith et al., 2023). A warming climate is expected to increase the likelihood of extreme high-temperature conditions, but changes in heatwave risk are not governed by shifts in mean temperature alone; they also depend on how variability, persistence, and the spatio-temporal organization of warm-season temperatures evolve (Barriopedro et al., 2023; Luo et al., 2024; Perkins-Kirkpatrick & Lewis, 2020). In many regions, the most disruptive events are those that persist for multiple days to weeks, because cumulative exposure limits physiological recovery, amplifies nighttime heat stress, and compounds risks through concurrent drought, wildfire potential, and air-quality degradation (Liu et al., 2025; Sutanto et al., 2020; Tripathy et al., 2023). These concerns are particularly salient for Canada, where rapid warming and increasing warm-season extremes are already straining public health and infrastructure planning, especially in densely populated southern corridors (Clark et al., 2025; Henderson et al., 2022). Despite substantial progress in heatwave detection and statistical projection, a persistent gap remains in translating projected changes in frequency, duration, and intensity into a physically interpretable understanding of why certain regions exhibit disproportionately strong amplification, especially in event persistence (Barriopedro et al., 2023; Bhattarai et al., 2025; Khodayar Pardo & Paredes-Fortuny, 2024; Perkins-Kirkpatrick & Lewis, 2020). Addressing this gap requires diagnostic approaches that connect event-based heatwave metrics to the underlying dynamical structures that can sustain prolonged exceedance conditions (Hochman et al., 2020; Ren et al., 2023).

A large body of work has therefore focused on quantifying heatwaves through event-based indices derived from daily temperature time series, typically using percentile exceedance thresholds and minimum-duration criteria to construct event catalogues and summarize frequency, duration, and intensity at gridded and regional scales (Keellings et al., 2018; Kirchmeier-Young et al., 2025; Perkins & Alexander, 2013). These approaches are well-suited for mapping where heatwave hazard increases and for comparing scenarios, but they remain largely descriptive: they do not directly reveal the dynamical mechanisms that organize warm-season variability and govern whether exceedance conditions dissipate quickly or persist for extended periods (Cowan et al., 2014; Fischer & Schär, 2010; Kirchmeier-Young et al., 2025). Mechanistic interpretations are often invoked indirectly, through associations with synoptic blocking, quasi-stationary circulation anomalies, land–atmosphere coupling, soil-moisture feedbacks, or regional advection regimes, but these links are challenging to establish consistently across large spatial domains and across climate projections, particularly when the goal is to attribute changes in event persistence rather than event occurrence alone (Adeyeri et al., 2025;

Barriopedro et al., 2023; Cai et al., 2024; Zhang et al., 2025). As a result, projections can robustly indicate that heatwaves become more frequent and intense while leaving open a central question for risk assessment: which components of the temperature field's multi-scale dynamics are most responsible for the emergence of long-duration exceedance regimes, and how do those components evolve under future forcing?

Recent studies have started to bridge this gap by adopting dynamical-systems and reduced-order representations that express climate fields as superpositions of coherent spatio-temporal structures, making dominant time scales and persistence more explicit than in purely event-statistical summaries (Ann et al., 2026; Çelik & Altunkaynak, 2024; Mankovich et al., 2025). A range of decomposition approaches has been explored for this purpose. Classical empirical orthogonal function / principal component decompositions (EOF/PCA) and their engineering analogues, such as proper orthogonal decomposition (POD), provide compact energy-ranked bases that are widely used to identify dominant spatial patterns and their temporal variability (Bauer-Marschallinger et al., 2013; Forootan et al., 2016; Volkov, 2014). However, because these bases are typically optimized for variance capture over an entire analysis window, they can mix multiple physical processes and time scales, and they do not directly encode characteristic frequencies or decay rates; consequently, connecting leading modes to event persistence or to the onset and termination of transient extremes often requires additional assumptions or post-processing (Mankovich et al., 2025; Zhang et al., 2014; Zhao et al., 2019). Extensions that emphasize spectral separation (such as wavelet-based multiscale analysis, singular spectrum analysis, and spectral/space–time POD) enable direct interrogation of variability across frequency bands and can better isolate low-frequency contributions relevant to persistence (Ghil et al., 2002; Towne et al., 2018). Yet these approaches may still face practical limitations for attribution of extremes: results can be sensitive to window choices, band definitions, or stationarity assumptions, and the extracted components do not necessarily provide a compact set of coherent structures whose activation can be tracked straightforwardly on heatwave days versus non-heatwave conditions (Ann et al., 2026; Mankovich et al., 2025; Schmidt & Colonius, 2020). In parallel, Koopman-inspired and operator-theoretic methods, including dynamic mode decomposition (DMD), offer a complementary lens by extracting modes with characteristic frequencies and decay rates, thereby providing interpretable diagnostics of oscillatory content and damping that relate directly to persistence mechanisms (Kutz, 2013; Manohar, 2018; Sato et al., 2024; Schmid, 2010). At the same time, standard DMD is commonly applied over fixed windows and can be sensitive to noise and non-stationarity; for climate fields, this can complicate interpretation when dynamical organization evolves within a season or when the goal is to separate slowly varying background structures from shorter-lived variability while preserving event-scale interpretability (Colbrook, 2023; Hemati et al., 2017; Kutz et al., 2016; Lapo et al., 2025).

Despite this methodological progress, applications that explicitly connect dynamical decompositions to heatwave event catalogues remain limited, and a clear need persists for workflows that link projected changes in heatwave frequency, duration, and intensity to the evolving dynamical structures that can sustain prolonged exceedance conditions and shape the spatial geography of heat-risk hotspots (Ann et al., 2026; Velagar et al., 2025).

Motivated by this gap, this study proposes and applies an integrated event–dynamical framework to quantify regional heatwave amplification and to interpret projected changes in terms of the multiscale organization of warm-season temperature variability. The approach combines a percentile-based heatwave event catalogue, used to compute standard frequency, persistence, and intensity metrics at grid-cell and subregional scales, with a multiresolution dynamic mode decomposition (mrDMD) of the raw daily mean temperature field to extract coherent modes organized by characteristic time scale and damping behavior. A cut-off sensitivity and reconstruction sufficiency analyses are used to select the retained low-frequency content relevant for persistence, after which two attribution diagnostics link the dynamical and event perspectives: a heatwave-conditioned participation ratio ranks modes by preferential activation during heatwave days, and a spatial alignment analysis evaluates whether the energy footprints of the heatwave-relevant modes coincide with persistence and intensity hotspots. The framework is demonstrated over southern Canada using high-resolution regional climate simulations under a high-emissions pathway for present-day, mid-century, and late-century periods. By jointly diagnosing what changes and why, the study provides process-informed evidence of persistence-driven heatwave regime shifts and offers transferable diagnostics to support model evaluation and climate-risk assessment.

## 2. Methodology

### 2.1 Framework overview

To systematically investigate how climate change alters the persistence and severity of regional heat risks, this study develops an integrated workflow (Fig. 1) that couples heatwave event diagnostics with a multiscale dynamical decomposition of warm-season temperature variability. The workflow is designed to translate projected changes in heatwave characteristics (i.e., frequency, duration, and intensity) into mechanistic insight by isolating the coherent dynamical structures that organize temperature anomalies and sustain exceedances. Heatwave amplification under future climate scenarios is therefore interpreted in terms of changes in the spatio-temporal organization of the warm-season temperature field, rather than summary statistics alone.

The first branch of the framework operates in an event-statistical setting. Heatwave days and events are identified using a percentile-based exceedance threshold combined with a minimum-duration constraint, producing an event catalogue from which standard metrics are computed. These metrics summarize occurrence (e.g., heatwave-day frequency and number of events), persistence (mean and maximum duration), and magnitude (intensity indices). They are first evaluated at the grid-cell level and then aggregated over user-defined subregions to support robust regional interpretation and inter-scenario comparison. In parallel, the second branch provides a dynamical representation of the temperature field using multiresolution dynamic mode decomposition (mrDMD). The primary analysis applies mrDMD directly to the daily mean temperature field, ensuring that the extracted modes represent the full warm-season signal used for the event diagnostics and impact interpretation. By separating variability across characteristic time scales, mrDMD isolates coherent spatio-temporal structures that can plausibly contribute to multi-day persistence and regional synchrony, features that are central to heatwave sustainment.

The final stage of the workflow integrates the event and dynamic branches through two attribution-oriented diagnostics. First, a heatwave-conditioned participation analysis uses the temporal modal coefficients to identify modes preferentially activated on heatwave days relative to non-heatwave days. Second, a spatial association analysis compares mode-energy (or amplitude) patterns with gridded maps of heatwave metrics to quantify alignment with persistence and intensity hotspots. Finally, reconstruction-based checks assess whether the retained modal representation reproduces the dominant warm-season variability relevant to heatwave behavior, providing a transparent sufficiency test that grounds the ensuing physical interpretation.

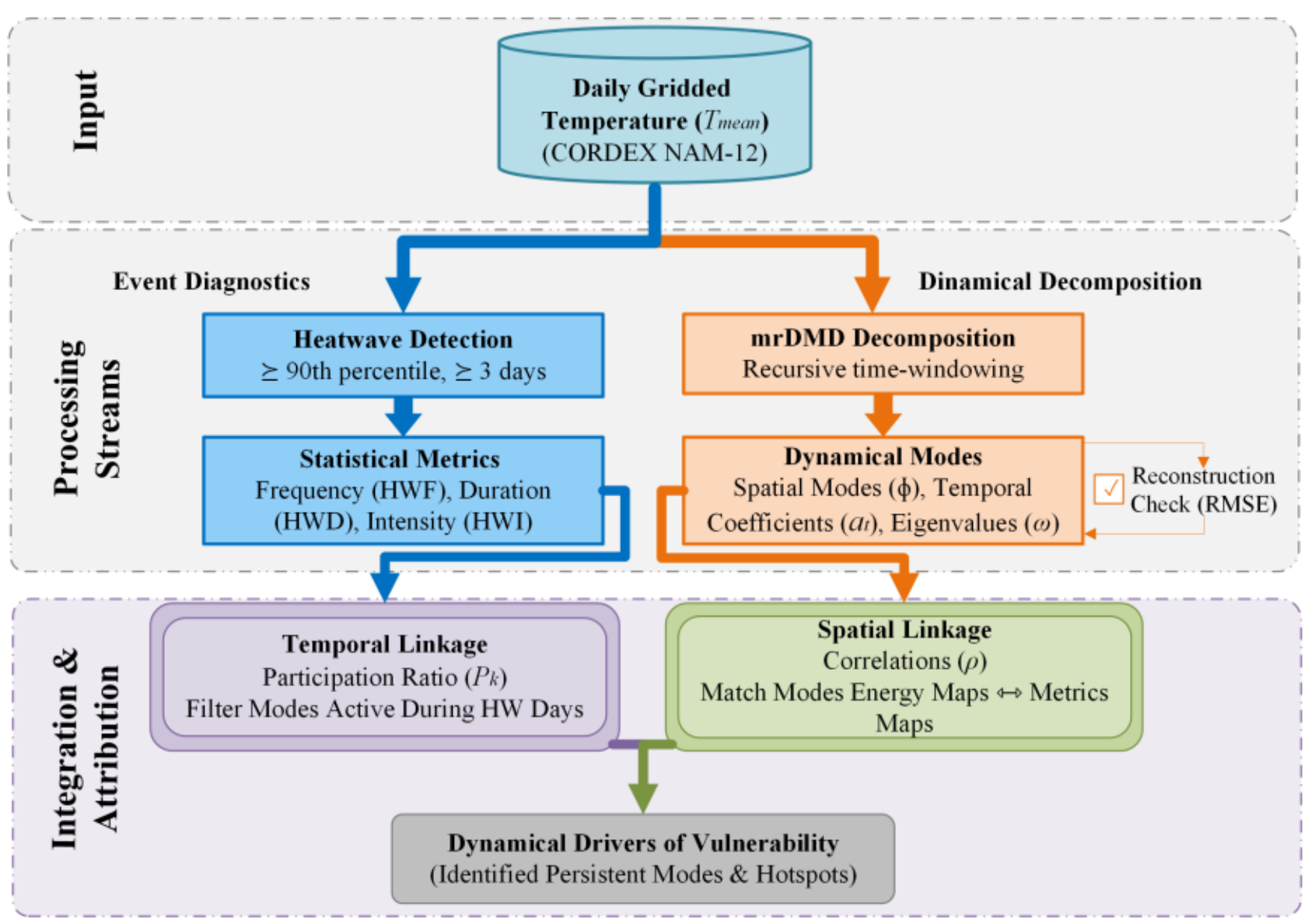

**Fig. 1** Schematic of the general analysis framework used in this study to relate heatwave diagnostics derived from gridded temperature data to a multiscale dynamical interpretation of the underlying temperature variability.

## 2.2 Heatwave identification and metrics

Heatwave events are identified at each grid cell $\boldsymbol{x}$ from the daily mean temperature time series $T(\boldsymbol{x}, t)$, where $t$ indexes days. For each calendar day (or day-of-season) $d$, a grid-cell-specific exceedance threshold $\theta_p(\boldsymbol{x}, t)$ is defined as the p-th percentile of temperatures in a reference climatology, estimated using a moving window centered on $d$ to ensure robust sampling. Let $d(t)$ map $t$ to its corresponding calendar day-of-season. A heatwave day is then defined through the exceedance indicator:

$$I(\boldsymbol{x}, t) = \begin{cases} 1, & if \quad T(\boldsymbol{x}, t) > \theta_p(\boldsymbol{x}, d(t)) \\ 0, & \text{otherwise} \end{cases} \tag{1}$$

Here $p$ is set to the 90th percentile. Heatwave events correspond to contiguous runs of exceedance days $\{I(\boldsymbol{x}, t) = 1\}$ with duration $L \geq L_{min}$ (e.g., $L_{min} = 3$ days). Event start and end dates are obtained by segmenting $I(\boldsymbol{x}, t)$ into consecutive sequences and retaining those meeting the minimum-duration criterion. Based on the resulting event catalogue, a set of heatwave metrics is computed over an analysis window $\mathcal{T}$ (e.g., a season) to summarize occurrence, persistence, and magnitude. Heatwave frequency (HWF) is defined as the total number of heatwave days:

$$\text{HWF}(\boldsymbol{x}) = \sum_{t \in \mathcal{T}} I(\boldsymbol{x}, t) \tag{2}$$

where $\mathcal{T}$ denotes the set of days in the analysis window. The Heatwave Number (HWN) is the count of retained heatwave events within $\mathcal{T}$. Persistence is quantified using event durations $L_j(\boldsymbol{x})$ (in days) for events $j = 1, \dots, \text{HWN}(x)$, including the Maximum Duration (HWD):

$$\text{HWD}(\boldsymbol{x}) = \max_j L_j(\boldsymbol{x}) \tag{3}$$

The Mean Duration (HWMD) is defined as:

$$\text{HWMD}(\boldsymbol{x}) = \frac{1}{\text{HWN}(x)} \sum_{j=1}^{\text{HWN}(x)} L_j(\boldsymbol{x}) \tag{4}$$

with the convention that HWMD = 0 when HWN = 0. Heatwave intensity (HWI) is summarized using an exceedance-based formulation computed on heatwave days, defined as the mean temperature anomaly above the threshold:

$$\text{HWI}(\boldsymbol{x}) = \frac{1}{\text{HWF}(x)} \sum_{t \in \mathcal{T}} I(\boldsymbol{x}, t) \left( T(\boldsymbol{x}, t) - \theta_p(\boldsymbol{x}, d(t)) \right) \tag{5}$$

with HWI = 0 when HWF = 0. In addition to HWI, the maximum heatwave magnitude (HWM) is defined here as the largest threshold exceedance attained on heatwave days within the analysis window:

$$\text{HWM}(\boldsymbol{x}) = max_{t \in \mathcal{T}} \left[ I(\boldsymbol{x}, t) \left( T(\boldsymbol{x}, t) - \theta_p(\boldsymbol{x}, d(t)) \right) \right] \tag{6}$$

with HWM = 0 when HWF = 0. Alternative intensity indices (e.g., event-mean exceedance, or cumulative exceedance) can be defined analogously, provided the temperature statistic used for intensity is consistent with that used to construct $\theta_p$. All metrics are first computed at the grid-cell scale to preserve spatial heterogeneity, and regional summaries are then obtained by area-weighted aggregation of grid-cell metrics over the defined subregions. These regional summaries describe regional heatwave exposure and are distinct from the regional heatwave-day definition adopted in the later participation analysis.

**2.3 Multiresolution dynamic mode decomposition**

Multiresolution dynamic mode decomposition (mrDMD) is used to represent the warm-season daily-mean temperature field as a superposition of coherent spatio-temporal structures, explicitly organized by characteristic time scales. The decomposition is applied to the raw daily mean temperature $T(\mathbf{x}, t_k)$ over the warm season to remain consistent with the event-based heatwave diagnostics described in Section 2.2, which are defined from exceedances of percentile-based thresholds in the original temperature field. Restricting the analysis to the warm season limits the influence of the annual cycle, while the multiresolution hierarchy separates slowly varying background variability from subseasonal structures most relevant to multi-day persistence. Let $\mathbf{x}_k \in \mathbb{R}^n$ denote the gridded temperature field at day $k (k = 1, \dots, m)$, where $n$ is the number of spatial grid points and $m$ is the number of daily samples in the analysis window. The snapshot sequence is arranged into a data matrix $\mathbf{X}$ as summarized in Eq. (7). From $\mathbf{X}$, two time-shifted matrices are formed from consecutive snapshots, $\mathbf{X}_1 = [\mathrm{x}_1, \dots, \mathrm{x}_{m-1}]$ and $\mathbf{X}_2 = [\mathrm{x}_2, \dots, \mathrm{x}_m]$, which defines the one-step evolution used in DMD:

$$\mathbf{x}_k = \mathrm{vec}\big(T(\mathbf{x}, t_k)\big) \in \mathbb{R}^n \tag{7}$$

Standard DMD approximates the best-fit linear operator $\mathbf{A}$ that advances the state by one sampling interval $\Delta t$ (with $\Delta t = 1$ day here) through $\mathbf{X}_2 \approx AX_1$. To improve numerical stability and filter noise, a truncated singular value decomposition is applied to $\mathbf{X}_1$, retaining rank $r = 20$ (selected to capture approximately 98% of the variance). The reduced operator is then constructed in the $r$-dimensional subspace, and eigen-decomposed to obtain DMD eigenvalues $\lambda_k$ and modes $\phi_k$. The corresponding temporal coefficients $b_k(t)$ provide a time-resolved representation of modal activity over the analysis window. Eigenvalues are interpreted in continuous time using $\omega_k = \ln(\lambda_k)/\Delta t$, where $\mathrm{Im}(\omega_k)$ characterizes the oscillation content, with associated period $P_C = 2\pi/|\mathrm{Im}(\omega_k)|$) and $\mathrm{Re}(\omega_k)$ controls exponential decay. Persistence is quantified through an e-folding time scale $\tau_k = 1/|\mathrm{Re}(\omega_k)|$ (days), so that weakly damped modes (large $\tau_k$) correspond to slowly varying, potentially heatwave-relevant structures, while strongly damped modes represent short-lived variability.

MrDMD extends DMD by repeating this decomposition over a hierarchy of nested time windows (schematically illustrated in Fig. 2). At level $\ell$, the analysis window is partitioned into $2^\ell$ non-overlapping segments, and DMD is performed independently within each segment. Coarser levels emphasize low-frequency, long-lived variability, whereas finer levels isolate shorter-lived dynamics. The maximum level is set to $L = 9$, yielding a shortest resolved window length of approximately $(m/2^L)\Delta t \approx 3$ days. In the mrDMD implementation, mode retention is controlled by a user-defined cut-off parameter $F_c$ reported in Table 1, that specifies the maximum number of oscillations permitted within a given time window. For a window of length $T_\ell$ at level $\ell$, the corresponding level-dependent cut-off is $F_\ell = F_c/T_\ell$. Modes are retained when their oscillatory content falls below this level-dependent threshold, so the retained set consists of modes that are slow relative to the time window at that level. A reconstruction-based sufficiency check is then used to verify that the retained mrDMD representation captures the dominant warm-season variability relevant to heatwave behavior. Reconstructions based on the retained modes are evaluated against the original temperature field using domain- and region-aggregated skill metrics, including RMSE, NRMSE, and $R^2$ over the analysis window (Chicco et al., 2021; Ranatunga et al., 2016). These diagnostics provide a transparent criterion for interpreting modal persistence and attribution results in subsequent sections.

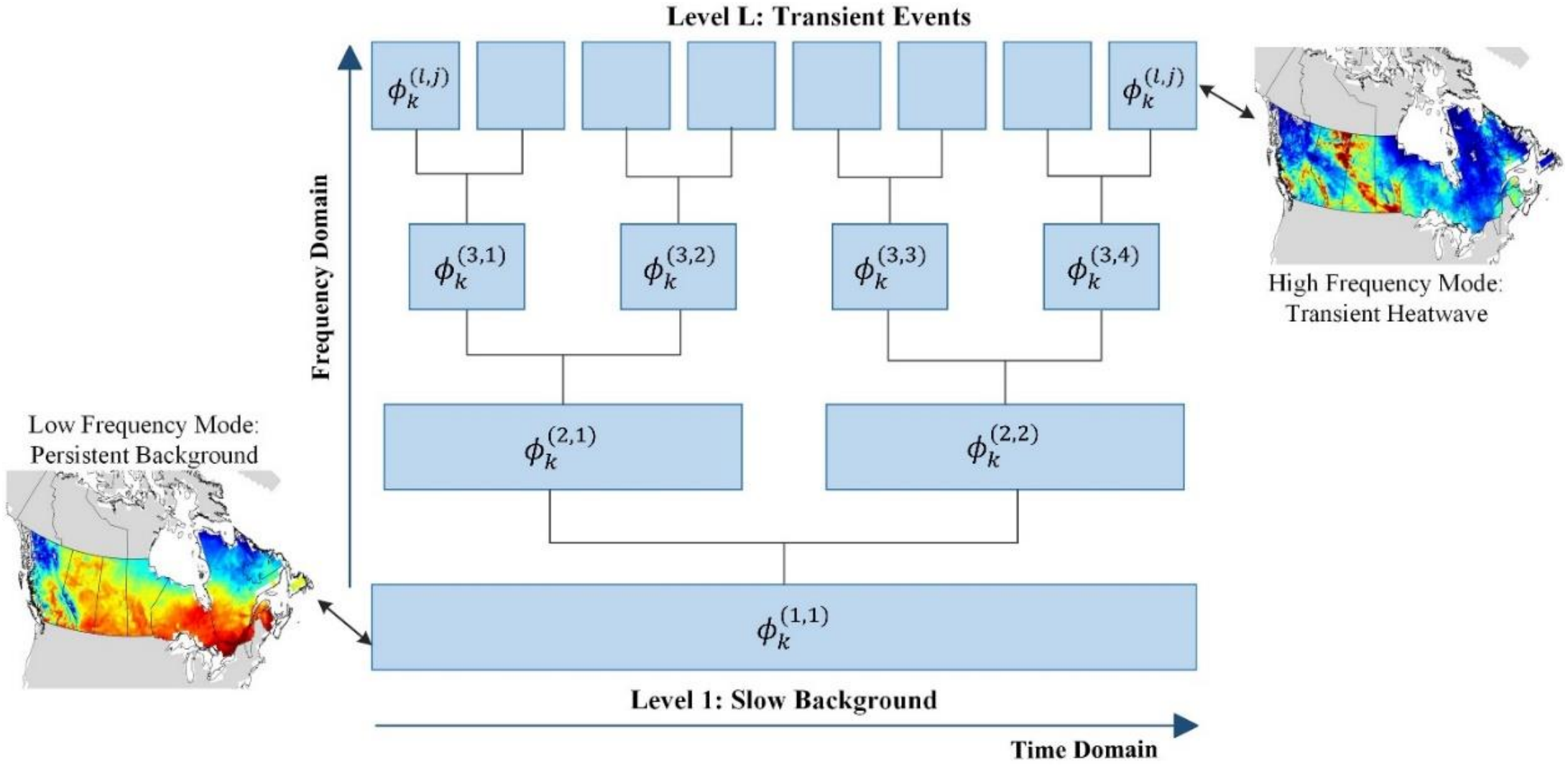

**Fig. 2** Schematic of the mrDMD framework, illustrating the hierarchical time-window partitioning and scale-separated extraction of dynamic modes across decomposition levels.

## 2.4 Integrated dynamical attribution

To identify which dynamical structures are most relevant to heatwave behavior, the mrDMD representation (Section 2.3) is coupled to the event-based classification of days into heatwave and non-heatwave conditions (Section 2.2). For each mrDMD mode $k$, the associated temporal coefficient $b_k(t)$ describes the time-varying contribution of that mode to the reconstructed temperature field. Because $b_k(t)$ is generally complex, a scalar amplitude is defined as $A_k(t) = |b_k(t)|$, providing a non-negative measure of modal activity that is independent of phase. Similarly, $\phi_k(x)$ denotes the spatial value of the mode $k$ at grid point $x$, and $|\phi_k(x)|$ represents its magnitude.

A mode participation ratio $P_k$ is then computed to quantify preferential activation during heatwave conditions. Let $\mathcal{H}$ and $\mathcal{N}$ denote, respectively, the sets of heatwave and non-heatwave days over the analysis window. For this binary regional classification, a regional heatwave day is defined by the area-weighted fraction of grid cells experiencing heatwave conditions within each subregion. A day is classified as a regional heatwave day when this fraction exceeds 50%; otherwise, it is classified as non-heatwave. Because the computation of $P_k$ requires a day-by-day classification of heatwave and non-heatwave conditions, this binary regional classification is introduced separately from the area-weighted regional summaries described above (Section 2.2), which represent aggregated exposure metrics rather than a day-by-day regional event mask. The participation ratio is then defined as the ratio of the mean modal amplitude on heatwave days to that on non-heatwave days:

$$P_k = \langle A_k(t) \rangle_{t \in \mathcal{H}} / \langle A_k(t) \rangle_{t \in \mathcal{N}} \quad (8)$$

where $\langle \, . \, \rangle$ denotes a temporal average. Values $P_k > 1$ indicate that mode $k$ is more active during heatwave days than during non-heatwave conditions, whereas $P_k \approx 1$ indicates comparable activation across both regimes. Modes are ranked by $P_k$, and the subset with the highest participation ratios is interpreted as the dominant set of heatwave-relevant dynamical structures for the considered region and climate period.

To evaluate whether these dominant modes also organize the spatial distribution of heatwave impacts, a spatial association analysis is conducted between gridded heatwave metrics and the composite footprint of the retained heatwave-relevant low-frequency modes. For each region and period, a composite dynamical footprint $E_{HW}(\mathbf{x})$ is constructed from the $top - k$ heatwave-relevant low-frequency modes identified by the participation analysis:

$$E_{\mathrm{HW}}(x) = \sum_{k \in \mathcal{K}} \omega_k |\phi_k(x)|^2 \quad (9)$$

where $\mathcal{K}$ denotes the retained heatwave-relevant modes and $\omega_k$ is a non-negative weight proportional to the heatwave-conditioned modal activity (e.g., $\omega_k \propto \langle |b_k(t)| \rangle_{t \in H}$ ). Throughout the main spatial-attribution analyses, $k = 5$. This construction emphasizes spatial patterns expressed by modes that are both low-frequency and preferentially activated during heatwave conditions. The correspondence between $E_{HW}(\mathbf{x})$ and the gridded heatwave-metric maps $M(\mathbf{x}) \in \{\mathrm{HWMD}(\mathbf{x}), \mathrm{HWI}(\mathbf{x})\}$ is quantified using two complementary measures: (i) an area-weighted Spearman rank correlation:

$$\rho_{HW,M} = \text{corr}_S\big(E_{HW}(\mathbf{x}), M(\mathbf{x})\big) \tag{10}$$

and (ii) a hotspot-overlap diagnostic based on the intersection of the upper-tail hotspot areas of $E_{HW}(\mathbf{x})$ and $M(\mathbf{x})$, where hotspots are defined as the top 20% of grid cells in each field. Large $|\rho_{HW,M}|$ values and strong hotspot overlap indicate that the composite dynamical footprint aligns closely with the geography of heatwave persistence and intensity hotspots.

## 3. Case Study

### 3.1 Case study description and data specification

The analysis focuses on the southern latitudes of Canada, where warm-season heat exposure and associated impacts are most concentrated. The study domain is partitioned into eight subregions for regional interpretation based on administrative boundaries: British Columbia (BC), Alberta (AB), Saskatchewan (SK), Manitoba (MB), Ontario (ON), Québec (QC), New Brunswick (NB), and Newfoundland and Labrador (NL). All computations are restricted to land grid cells; ocean points are excluded using the land–sea mask dataset (Strandgren, 2023). The spatial extent of the domain and the adopted subregional partitioning are shown in Fig. 3.

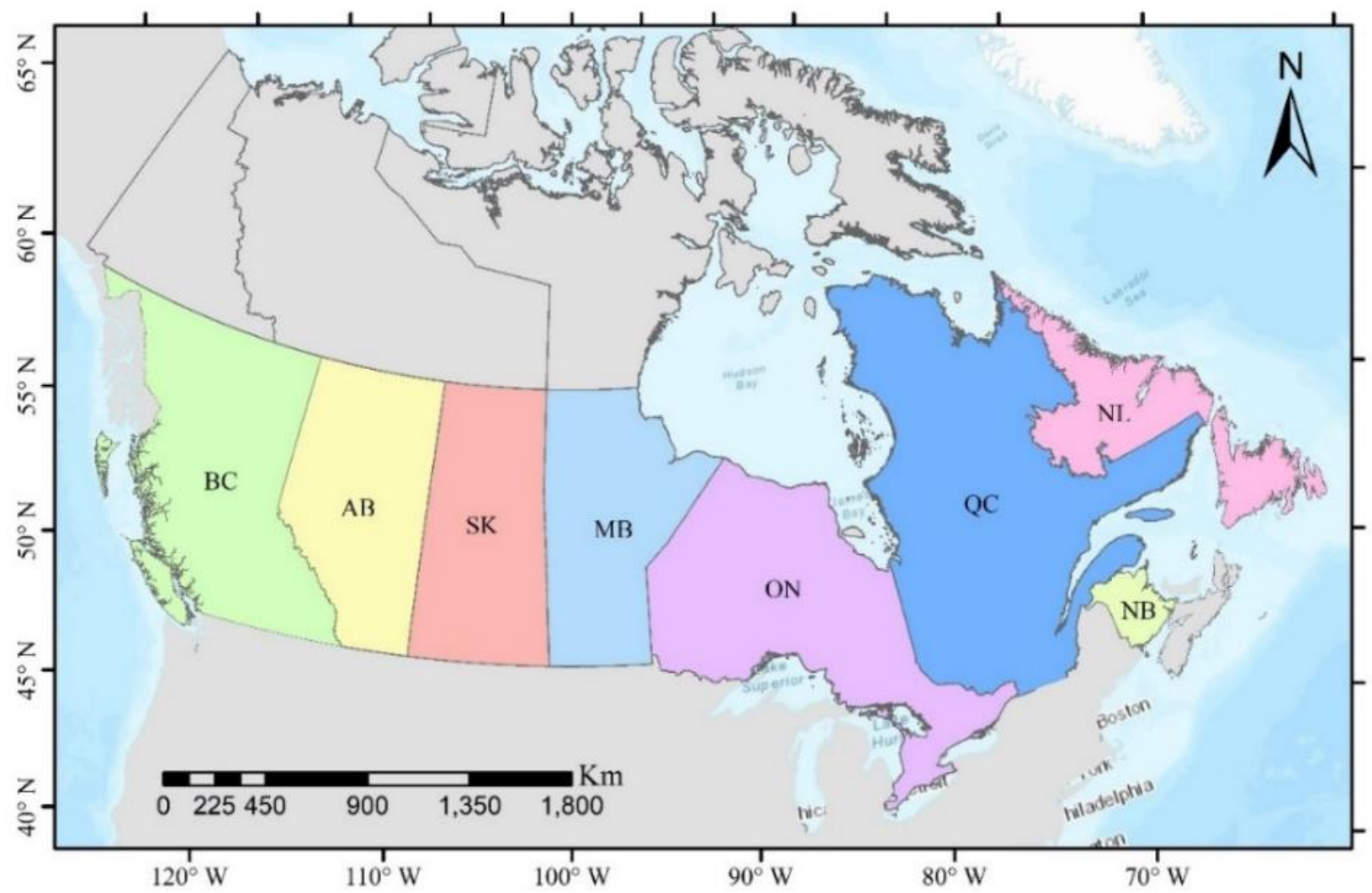

**Fig. 3** Study domain and regional partitioning over southern Canada.

Daily mean near-surface air temperature $T_{mean}$ is taken from the CORDEX North America (NAM-12) regional climate simulations, using the OURANOS CRCM5 dynamical downscaling of CanESM5 under the SSP5–8.5 scenario (*CanESM5 Hourly Dataset*). As the source archive provides hourly near-surface air temperature fields, daily $T_{mean}$ was computed by temporally averaging the hourly values for each day prior to the analyses. The resulting fields are analyzed on a regular latitude–longitude grid at 0.11° (≈12 km) horizontal resolution. After land masking, the analysis includes $n = 36{,}163$ land grid cells over the full domain. The analysis is restricted to the extended warm season (May–September) to capture the full life cycle of heatwave activity. Heatwave characteristics are evaluated over three 10-year time slices: a present-day period (2016–2025), a mid-century period (2051–2060), and a late-century period (2091–2100). Ten-year windows provide a practical balance between sampling of sub-seasonal extremes and limiting non-stationarity within each analysis block.

Heatwaves are defined using a fixed, grid-cell-specific percentile threshold derived from a historical reference climatology (2001–2010) (*CanESM5 Hourly Dataset*). For each grid cell and calendar day within May–September, the exceedance threshold $\theta_p(x,t)$ is estimated as the $p$-th percentile with $p = 90$, using a centered moving window of width $\omega = 15$ days to increase sampling robustness while preserving the seasonal progression. This window length was selected to provide sufficient sample size for robust estimation of calendar-day percentile thresholds while retaining seasonal specificity, consistent with established percentile-based heatwave methodologies (Barriopedro et al., 2023; Perkins & Alexander, 2013; Perkins-Kirkpatrick & Lewis, 2020). The threshold field is then held fixed across all periods, so that projected changes quantify exceedances relative to a consistent historical baseline.

Heatwave events are identified using a minimum duration criterion of $L_{min} = 3$ consecutive days above $\theta_p(x,t)$. Grid-cell metrics are subsequently aggregated within each subregion using area-weighted averaging to obtain regional summaries consistent with the reporting in Section 3.2. For the mrDMD analysis, the maximum decomposition level is set to $L = 9$ to resolve daily-to-subseasonal variability, and the truncation rank is set to $r = 20$ (where $r$ denotes the number of singular modes retained at each local decomposition in the reduced-order mrDMD representation) selected so that the cumulative singular-value energy captured from the corresponding snapshot matrix is approximately 98%, while suppressing noise. Table 1 summarizes the dataset, analysis periods, and key parameter settings.

**Table 1** Summary of Experimental Design and Parameters

| Category | Parameter | Setting |
|---|---|---|
| **Dataset** | Model Chain | CORDEX NAM-12 (OURANOS CRCM5 / CanESM5) |
| | Scenario | SSP5-8.5 (High Emissions) |
| | Variable | Daily Mean Temperature ($T_{mean}$) |
| | Resolution | 0.11° (~12 km) |
| **Time Domain** | Time period | May – September |
| | Present-Day Period | 2016 – 2025 |
| | Mid-Century Period | 2051 – 2060 |
| | Late-Century Period | 2091 – 2100 |
| **Heatwave Definition** | Threshold | 90th Percentile |
| | Reference Period | 2001 – 2010 (Fixed Baseline) |
| | Window | 15-day centered moving window |
| | Min. Duration ($L_{min}$) | ≥ 3 consecutive days |
| **mrDMD Settings** | Max. Levels ($L$) | 9 |
| | Decomposition Input | Raw temperature |
| | Truncation rank ($r$) | 20 |
| | mrDMD cut-off frequency ($F_c$) | 2 |

### 3.2 Heatwave metrics

This subsection summarizes the projected evolution of heatwave occurrence, persistence, and magnitude across the eight subregions using threshold-based event diagnostics. All metrics are computed at the grid-cell level for each May–September season and then aggregated regionally to facilitate inter-regional comparison. Spatial maps are presented using consistent colour scales across periods to enable direct visual comparison.

#### 3.2.1 Spatial patterns

Figure 4 summarizes the gridded evolution of heatwave characteristics across southern Canada for the present-day period (2016–2025), mid-century (2051–2060), and late-century (2091–2100). For clarity, Fig. 4 presents only the spatial distributions of HWF, HWMD, and HWI; the complementary persistence and peak-intensity metrics HWD and HWM are discussed through regional summaries in Section 3.2.2 and Table 2 rather than mapped here. The heatwave-frequency maps (HWF; heatwave days per May–September season) indicate a clear intensification and spatial expansion of exceedance conditions over time. During the present-day period, exceedances of the fixed historical percentile threshold remain spatially heterogeneous, with the largest HWF values concentrated over continental interiors. By mid-century, HWF increases across nearly all land grid cells, with a domain-wide shift toward more frequent heatwave days. Across the southern Canada domain, typical values rise from about 7–8 days/season in the present-day period to 38–39 days/season in the mid-century period. Late-century conditions exhibit widespread exceedance days across the domain and the emergence of pronounced frequency hotspots over the Prairie provinces, with SK showing peak HWF values of about 97 days/season.

The strongest qualitative change is observed in the persistence-related behaviour, clearly reflected in the mapped HWMD field and further supported by the regional HWD summaries reported in Section 3.2.2. In the present-day period, heatwave events are typically limited to short episodes (e.g., $\mathrm{HWMD} \approx 3.49$ days/event; $\mathrm{HWD} \approx 4$ days), whereas late-century maps show the development of multi-week persistence hotspots in several interior regions (e.g., HWD increasing up to 69 days in SK/MB and adjacent areas). This spatial transition indicates that projected changes are not limited to how often heatwave conditions occur, but also involve a marked increase in how long exceedance conditions persist once established, consistent with a shift toward warm-season regimes dominated by extended exceedance sequences rather than isolated extremes.

Intensity metrics also strengthen across the domain. The mapped mean intensity (HWI) also exhibits systematic increases from the present-day to the late century, with the largest intensification concentrated over the continental interior (e.g., SK/MB HWI increasing from 2.26 to 6.88 $K$ above threshold; HWM reaching 15.65 $K$), while the regional HWM summaries confirm that peak exceedances intensify as well. In contrast, coastal and maritime-influenced regions exhibit comparatively weaker intensification and/or shorter persistence; for example, late-century HWMD reaches 26.66 and 23.46 days/event in SK and MB, compared with 22.56 in BC, 17.68 in NB, and 16.01 in NL, while HWI reaches 6.88 and 6.80 K in SK and MB versus 5.81 K in BC, 4.48 K in NB, and 6.54 K in NL. Overall, the spatial patterns in Fig. 4 demonstrate that late-century amplification is characterized by the co-emergence of higher exceedance magnitudes and substantially longer event persistence, motivating the regional diagnostics in Section 3.2.2 where changes in event counts and durations are explicitly quantified.

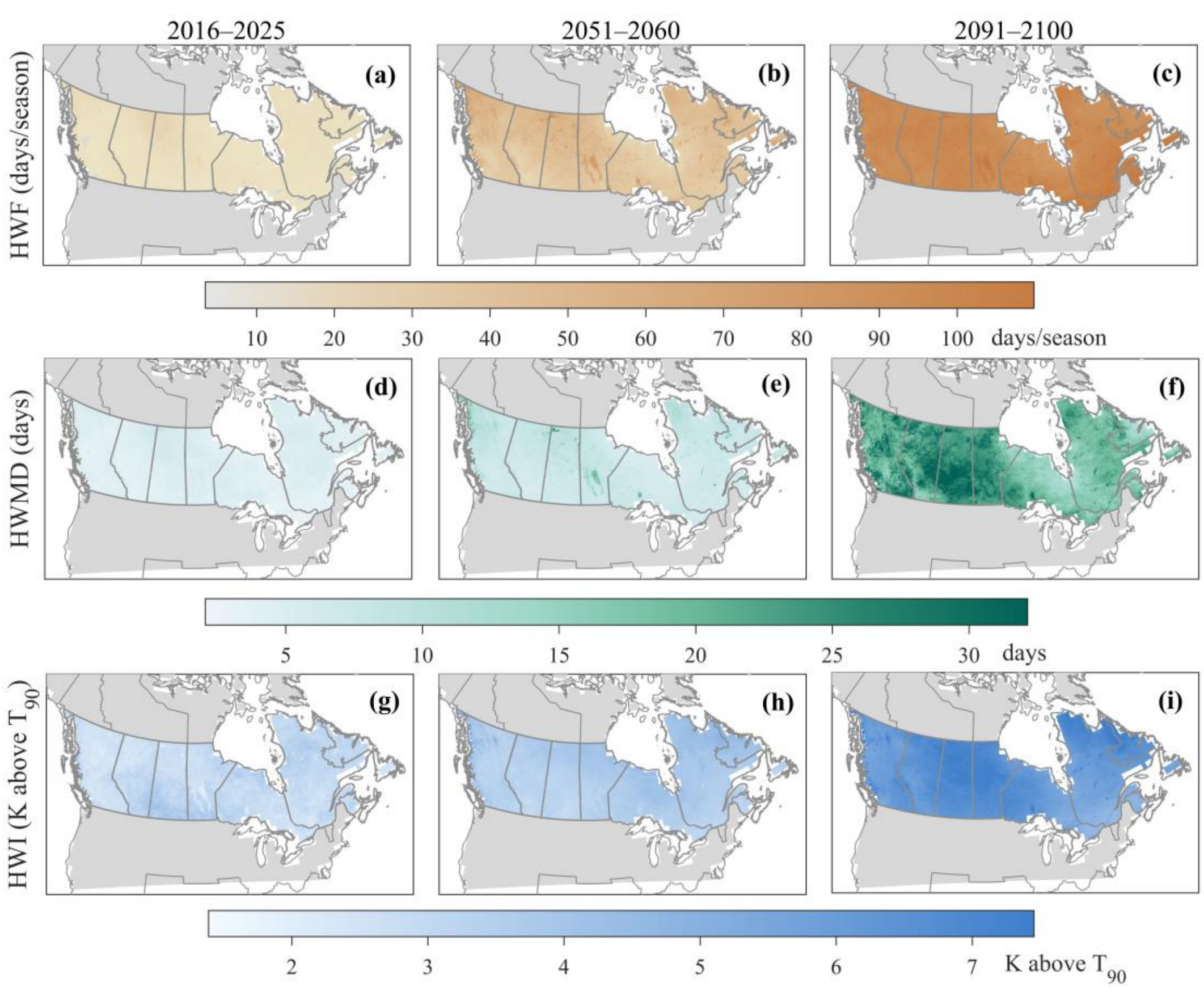

**Fig. 4** Spatial evolution of heatwave characteristics over southern Canada under SSP5–8.5. Maps show (a–c) heatwave frequency (HWF; heatwave days over May–September season), (d–f) mean heatwave duration (HWMD; days), and (g–i) mean heatwave intensity (HWI; K above $T_{90}$) for 2016–2025, 2051–2060, and 2091–2100.

### 3.2.2 Regional evolution of frequency, persistence, and intensity

To complement the gridded patterns in Section 3.2.1, heatwave characteristics are summarized at the subregional scale by aggregating grid-cell metrics over BC, AB, SK, MB, ON, QC, NB, and NL (area-weighted). Results are reported for the present-day (2016–2025), mid-century (2051–2060), and late century (2091–2100) periods using the fixed historical threshold defined in Section 3.1. Regional frequency, persistence, and intensity metrics are summarized in Table 2 for the present-day, mid-century, and late-century periods.

Across all regions, the aggregated metrics indicate strong amplification in regional heatwave exposure, with pronounced geographical contrasts. At present-day, regional HWF spans approximately 5–11 days per warm season, increasing to 90–97 days by the late century. At the same time, the regional summaries of HWN and HWMD also increase, indicating that the exposure amplification is accompanied by more frequent and more persistent heatwave conditions at the grid-cell level. Figure 5a summarizes the regional variation in mean event duration and highlights a systematic shift toward longer heatwaves across subregions from present-day to late century, with the strongest duration amplification concentrated over the continental interior (late-century HWMD reaching 26.66 days/event in SK). Figure 5b further clarifies the nature of this evolution by mapping regional centroids in (HWN, HWMD) space. The displacement of centroids across periods indicates that late-century amplification cannot be explained by event counts alone. Several regions exhibit a marked persistence-driven shift (substantial increase in HWMD with a comparatively smaller increase in HWN), whereas others show a more balanced diagonal shift in which both the number of events and their typical duration increase. This diagnostic therefore supports the interpretation that the heatwave regime evolves toward more persistent exceedance sequences in key interior regions under SSP5–8.5, consistent with the persistence hotspots identified in Section 3.2.1.

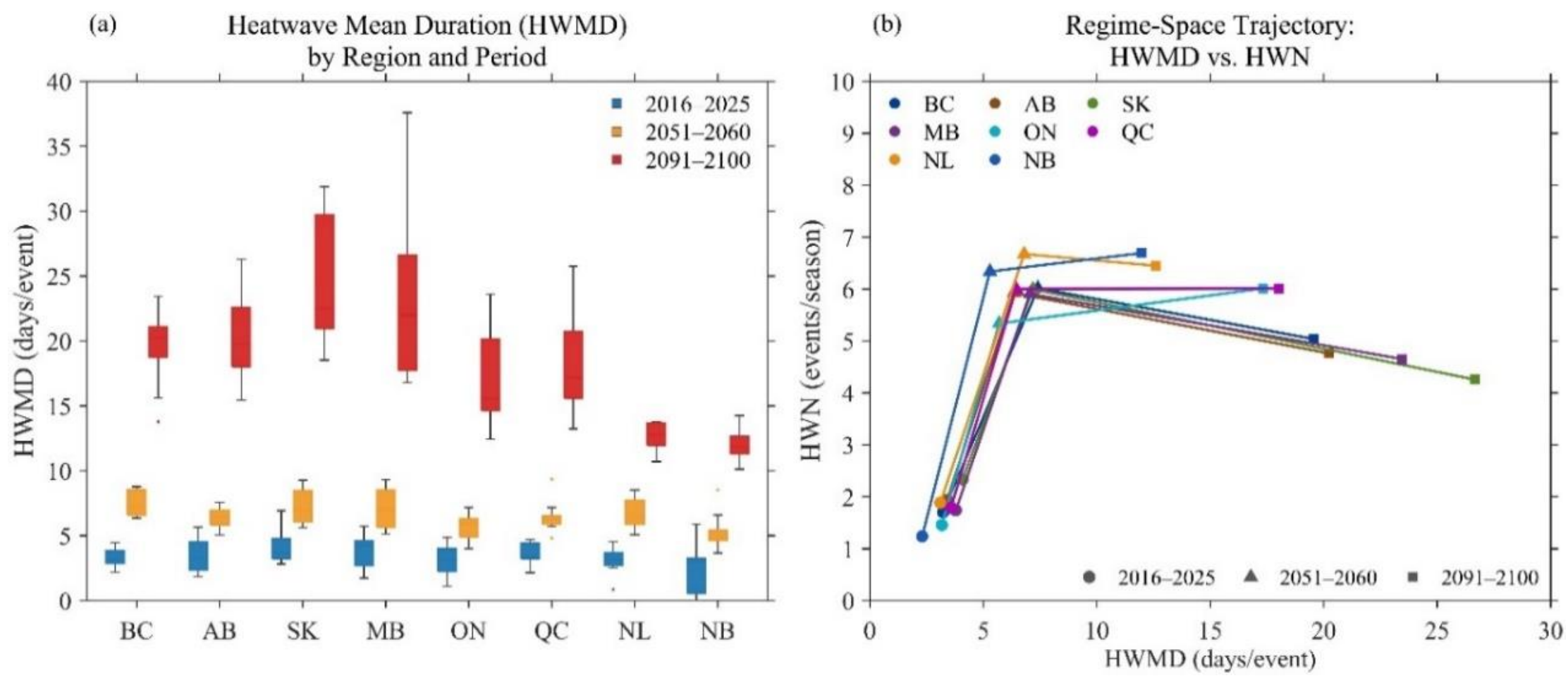

**Fig. 5** Regional duration amplification and heatwave regime shift under SSP5–8.5. (a) Distribution of mean heatwave duration (HWMD; days/event) across the eight subregions for 2016–2025, 2051–2060, and 2091–2100. (b) Regional centroids in (HWN, HWMD) space (HWN: events/season; HWMD: days/event), shown with distinct markers for each period.

Intensity amplification accompanies the frequency–persistence changes across all regions. Regional mean intensity (HWI) increases from present-day values of approximately 1.7-2.4 $K$ above threshold to late-century values of 4.5-6.9 $K$, with the largest increases concentrated over the continental interior; for example, late-century HWI reaches 6.88 K in SK and 6.80 K in MB (Table 2). Peak exceedances (HWM) also intensify, with late-century regional maxima reaching 15.65 $K$ above threshold. Collectively, these results indicate that future heat risk is shaped simultaneously by more frequent exceedance, greater persistence, and higher above-threshold magnitude, motivating the dynamical attribution analyses in Sections 3.3–3.4 to assess whether projected persistence amplification is associated with changes in the dominant low-frequency structures of warm-season temperature variability.

**Table 2** Regional heatwave frequency, persistence, and intensity metrics under SSP5–8.5.

| Region | Period | HWF (days/season) | HWN (events/season) | HWMD (days/event) | HWD (days) | HWI (K) | HWM (K) |
|---|---|---|---|---|---|---|---|
| BC | 2016-2025 | 7 | 2 | 3.23 | 4 | 1.76 | 3.49 |
|  | 2051-2060 | 42 | 6 | 7.41 | 16 | 2.89 | 7.42 |
|  | 2091-2100 | 95 | 6 | 22.56 | 60 | 5.81 | 13.42 |
| AB | 2016-2025 | 8 | 2 | 3.42 | 4 | 1.98 | 4.20 |
|  | 2051-2060 | 38 | 6 | 6.35 | 13 | 2.72 | 6.91 |
|  | 2091-2100 | 95 | 5 | 24.22 | 63 | 6.35 | 14.64 |
| SK | 2016-2025 | 11 | 2 | 4.09 | 5 | 2.26 | 4.92 |
|  | 2051-2060 | 42 | 6 | 7.19 | 15 | 2.73 | 7.23 |
|  | 2091-2100 | 97 | 5 | 26.66 | 69 | 6.88 | 15.65 |
| MB | 2016-2025 | 8 | 2 | 3.79 | 4 | 2.11 | 4.52 |
|  | 2051-2060 | 41 | 6 | 7.06 | 14 | 3.05 | 7.76 |

| | 2091-2100 | 94 | 5 | 23.46 | 69 | 6.80 | 15.55 |
|---|---|---|---|---|---|---|---|
| ON | 2016-2025 | 6 | 1 | 3.17 | 4 | 2.14 | 4.26 |
| | 2051-2060 | 31 | 5 | 5.70 | 10 | 3.11 | 7.83 |
| | 2091-2100 | 92 | 6 | 17.33 | 56 | 5.80 | 14.34 |
| QC | 2016-2025 | 8 | 2 | 3.58 | 4 | 2.22 | 4.60 |
| | 2051-2060 | 38 | 6 | 6.44 | 13 | 3.25 | 8.85 |
| | 2091-2100 | 97 | 6 | 18.03 | 58 | 5.79 | 15.16 |
| NL | 2016-2025 | 7 | 2 | 3.11 | 4 | 2.21 | 4.47 |
| | 2051-2060 | 41 | 7 | 6.80 | 14 | 3.52 | 9.86 |
| | 2091-2100 | 92 | 7 | 16.01 | 48 | 6.54 | 14.12 |
| NB | 2016-2025 | 5 | 1 | 2.31 | 3 | 1.81 | 3.64 |
| | 2051-2060 | 33 | 6 | 5.29 | 9 | 2.57 | 6.93 |
| | 2091-2100 | 90 | 7 | 17.68 | 51 | 4.48 | 12.05 |

### 3.3 mrDMD analysis of temperature variability

#### 3.3.1 Selection of the dynamical content retained for attribution

The attribution analyses that follow rely on a reduced mrDMD representation that retains the dominant components of warm-season temperature variability while filtering variability that is either too short-lived or too weak to support persistent exceedance conditions. A sensitivity analysis is therefore performed to select the retained dynamical content (cut-off parameter and associated mode set) and to verify that this choice provides a faithful reconstruction of the warm-season temperature field. The selection is guided by two criteria: (i) interpretability for persistence, favoring low-frequency structures expected to sustain multi-day heatwave conditions, and (ii) reconstruction sufficiency, ensuring that the retained modes reproduce the dominant variability relevant to heatwave diagnostics.

Candidate mode sets are constructed by applying a mrDMD cut-off parameter $F_c$ defined in Section 2.3. Five candidate cut-offs are evaluated, $F_c = 1$, 1.5, 2, 2.5, and 3. For each candidate cut-off, the retained-mode reconstruction $\hat{T}(x,t)$ is formed and compared against the original warm-season temperature field $T(x,t)$ over the analysis window.

Reconstruction sensitivity to the five candidate cut-offs is summarized in Fig. 6a using time-averaged, domain-wide skill metrics. The results indicate that reconstruction skill varies systematically with the selected cut-off parameter. In particular, increasing $F_c$ from 1 to 2 reduces RMSE from 1.65 to 1.51 K and improves $R^2$ . Beyond $F_c = 2$, reconstruction skill deteriorates slightly, with RMSE increasing again for $F_c = 2.5$ and $F_c = 3$. Similar behavior is observed for region-aggregated time-averaged metrics, indicating that the selected representation maintains consistent fidelity across subregions (regional RMSE range: 1.45 to1.61 $K$).

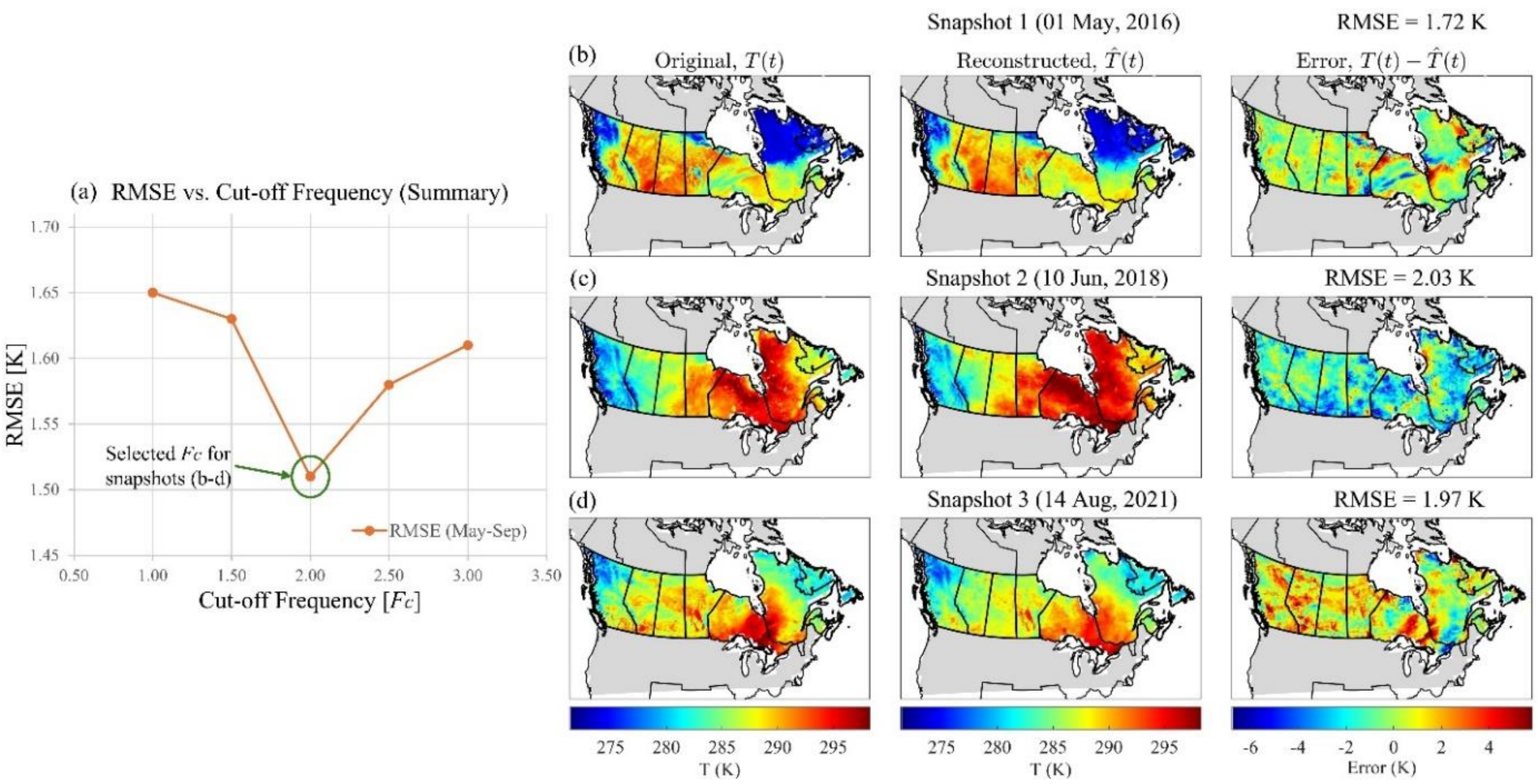

**Fig. 6** Sensitivity of reconstruction skill to the mrDMD cut-off parameter. (a) Time-averaged domain-wide reconstruction RMSE for five cut-off parameters, and (b-d) illustrative snapshots comparing the original temperature field, the reconstructed field using the selected cut-off, and the corresponding error maps over southern Canada (land-only).

The retained dynamical content is also supported by the distribution of modal amplitude across mrDMD levels and periods. Fig. 7 (mrDMD level–period amplitude maps) highlights a progressive redistribution of dominant activity toward lower-frequency levels in the continental interior (e.g., SK) relative to coastal and maritime regions (e.g., BC and NL), consistent with the emergence of stronger persistence hotspots in Section 3.2. This behavior motivates an emphasis on the low-frequency structures most plausibly associated with multi-day persistence.

Based on the sensitivity analysis in Fig. 6a and the dynamical evidence in Fig. 7, the intermediate cut-off $F_c = 2$ is adopted for the subsequent attribution results because it provides a robust compromise between reconstruction sufficiency and interpretability: it targets the range of time scales where the redistribution toward low-frequency activity is most apparent while maintaining high reconstruction skill. To contextualize the choice of mrDMD, we compared its reconstruction skill with POD and standard DMD under the same analysis window and retained-rank setting. For the selected domain, mrDMD yielded the best overall reconstruction fidelity (time-averaged RMSE = 1.51 K, NRMSE = 3 %, $R^2$ = 0.94), compared with POD (1.92 K, 4 %, 0.91) and DMD (2.18 K, 5 %, 0.89), supporting its use for the subsequent attribution analysis. Illustrative warm-season snapshots comparing $T(x,t)$, $\hat{T}(x,t)$, and the residual $T - \hat{T}$ for the selected cut-off are shown in Fig. 6b–d, providing a visual sufficiency check over land grid cells and confirming that the retained modes reproduce the dominant spatial organization relevant to the ensuing attribution analyses. The chosen cut-off and retained-mode configuration are applied consistently across regions and periods to enable physically meaningful comparisons of modal persistence, heatwave-conditioned activation, and spatial alignment in the following sections.

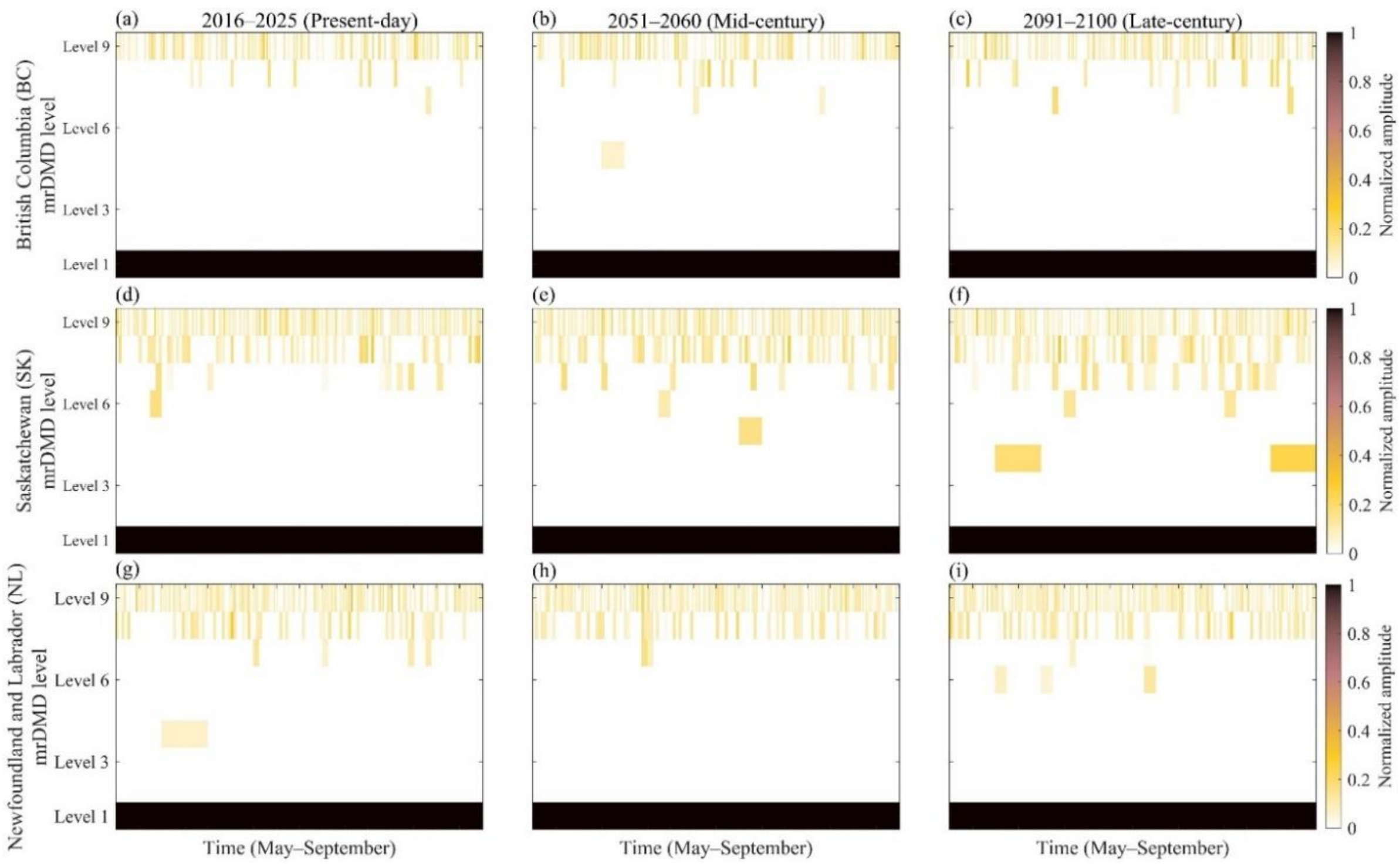

**Fig. 7** mrDMD time–frequency amplitude maps across regions and periods. Checkerboard maps show the distribution of modal amplitude across mrDMD levels for BC, SK, and NL during 2016–2025, 2051–2060, and 2091–2100, highlighting a progressive redistribution of dominant activity toward lower-frequency levels in the interior relative to coastal regions.

### 3.3.2 Modal damping and persistence under climate change

The retained mrDMD representation in Section 3.3.1 enables a direct characterization of warm-season persistence through the damping properties of the dominant modes. Here, damping refers to the rate at which the amplitude of a dynamical mode decays in time: strongly damped modes decay rapidly and therefore represent short-lived variability, whereas weakly damped modes persist longer and can support sustained warm anomalies relevant to heatwave duration. For each retained mode $k$, persistence is quantified by the real part of the continuous-time eigenvalue $\omega_k$ (Section 2.3). Specifically, $\mathrm{Re}(\omega_k)$ controls exponential decay and is summarized via an e-folding time scale $\tau_k = 1/|\mathrm{Re}(\omega_k)|$ (days), such that larger $\tau_k$ indicates weaker damping and a greater propensity for sustained anomalies. This persistence perspective is directly relevant to the duration amplification reported in Section 3.2, where changes in heat risk are expressed most strongly through longer event persistence.

The persistence shift is also evident in the complex-plane distributions of the dominant mrDMD eigenvalues (Fig. 8). Relative to the present-day, late-century eigenvalues in interior regions shift toward the weak-damping limit, consistent with reduced decay rates and longer e-folding time scales. In contrast, coastal and maritime regions exhibit comparatively smaller shifts in the eigenvalue cloud, aligning with their weaker duration amplification. This complex-plane diagnostic therefore provides an independent dynamical signature supporting the emergence of a more persistence-dominated warm-season regime in the continental interior.

Consistent with Fig. 8, the distribution of $\tau_k$ shifts upward by mid-century and further by late century, with the largest changes occurring over interior regions associated with strong persistence amplification (e.g., SK/MB). For example, the median e-folding time increases from 12 days (present-day) to 23 days (late century) in the Prairie interior, whereas representative coastal/maritime regions (e.g., BC/NL) exhibit smaller changes (median $\tau_k$ from 15 to 19 days). Complementarily, the fraction of retained modes exceeding a persistence threshold $\tau_k \geq \tau_0$ (with $\tau_0 = 10$ days) increases from 92% to 95% in the interior, indicating an expanding set of slowly decaying structures capable of supporting multi-day exceedance conditions.

To connect these damping changes to the scale structure emphasized by mrDMD, persistence is also summarized by decomposition level. Coarser levels (longer window lengths and lower characteristic frequencies) exhibit systematically larger $\tau_k$ than finer levels, as expected, but late-century changes are not uniform across levels. Instead, interior regions show a disproportionate strengthening of persistence at the lower-frequency levels retained by the selected cut-off $F_c$, consistent with the redistribution of modal amplitude toward these levels noted in Fig. 7. This pattern supports the interpretation that late-century duration amplification is associated not only with more frequent exceedance, but also with a dynamical reweighting toward slow, weakly damped structures that sustain persistent warm-season temperature anomalies.

These damping diagnostics provide a compact dynamical explanation for the regime shift observed in Fig. 5: as the warm-season dynamics become increasingly dominated by weakly damped low-frequency structures, exceedance sequences are more likely to persist once initiated, increasing HWMD and HWD without requiring a proportional increase in event counts. The subsequent attribution results, therefore, focus on the subset of retained modes with both high heatwave-conditioned participation and weak damping, and evaluate whether their spatial energy patterns align with regional hotspots of persistence and intensity.

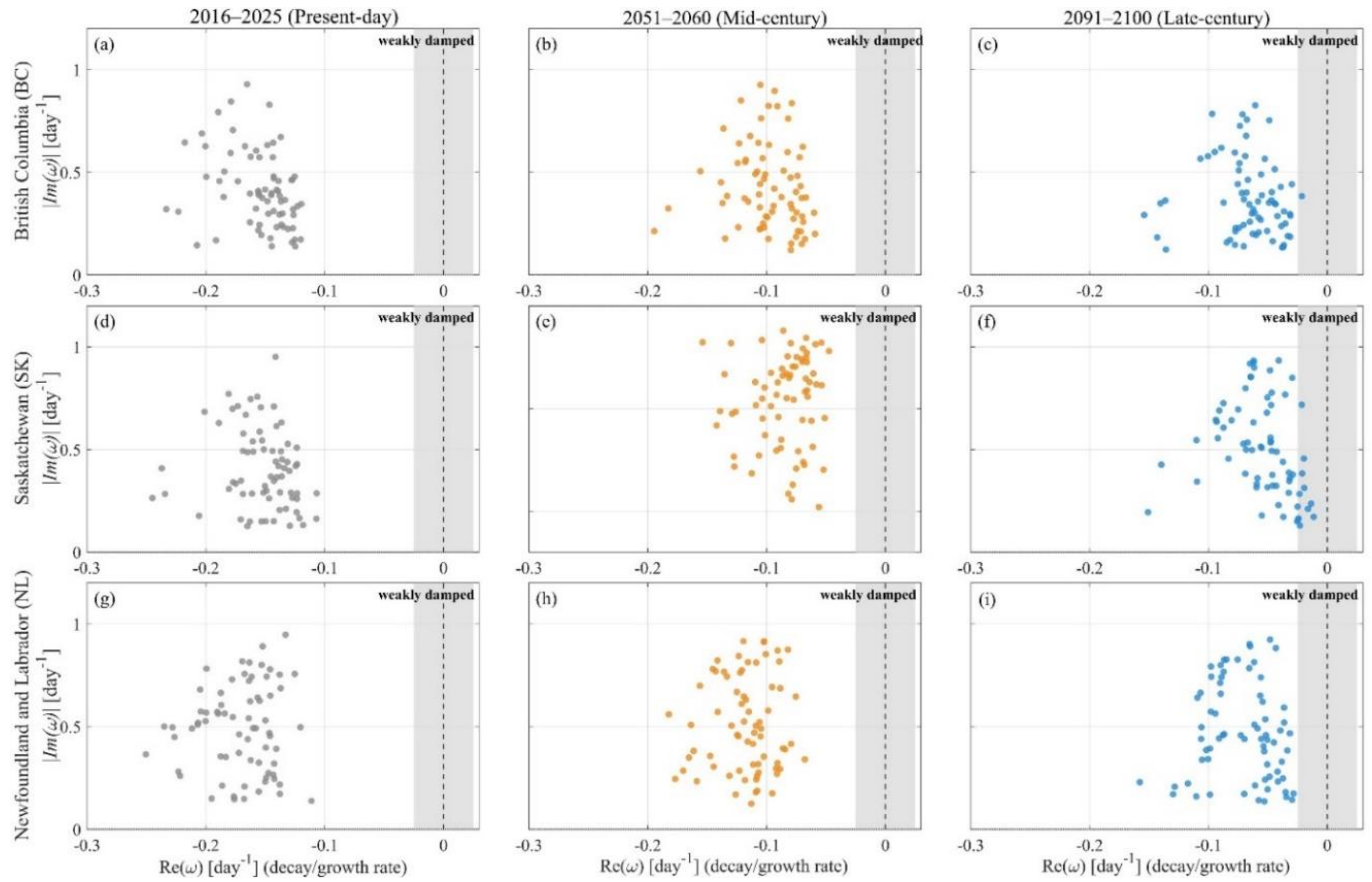

**Fig. 8** Complex-plane distributions of dominant mrDMD eigenvalues for British Columbia, Saskatchewan, and Newfoundland and Labrador across 2016–2025, 2051–2060, and 2091–2100.

### 3.3.3 Spatial structure of dominant modes

Having established that late-century dynamics exhibit weaker damping and a redistribution of activity toward low-frequency levels (Sections 3.3.1–3.3.2), the spatial structure of the dominant retained modes is examined to assess whether persistence amplification is accompanied by increased spatial coherence. This analysis focuses on the leading low-frequency mrDMD mode (i.e., the mode with the largest contribution within the retained low-frequency band under the selected cut-off $F_c$), because such modes are the most plausible candidates for organizing sustained, spatially extensive warm-season anomalies.

Figure 9 maps the normalized spatial amplitude $|\phi_k|$ of this leading low-frequency mode for BC, SK and NL across the present-day, mid-century, and late-century periods. Two contrasts are apparent. First, the interior region (SK) exhibits a progressive strengthening and spatial consolidation of the dominant low-frequency pattern: relative to the present-day, the late-century mode displays higher-amplitude contiguous structures and reduced small-scale fragmentation, consistent with an increasingly coherent, large-scale background conducive to multi-day persistence. Second, the coastal/maritime region (NL) exhibits comparatively weaker changes in spatial organization across periods, with the dominant mode remaining more spatially diffuse and/or fragmented. These differences align with the regional contrasts in duration amplification (Section 3.2.2) and the weaker interior-to-coastal shift in eigenvalue damping identified in Fig. 8.

Overall, Fig. 9 indicates that persistence amplification in interior regions is accompanied by an increasingly coherent low-frequency spatial structure, supporting the interpretation that late-century heatwave regimes are shaped by slowly varying, spatially organized background anomalies. This motivates the integrated attribution analysis in Section 3.4, which evaluates whether modes exhibiting high heatwave-conditioned participation also align spatially with regional patterns of heatwave persistence and intensity.

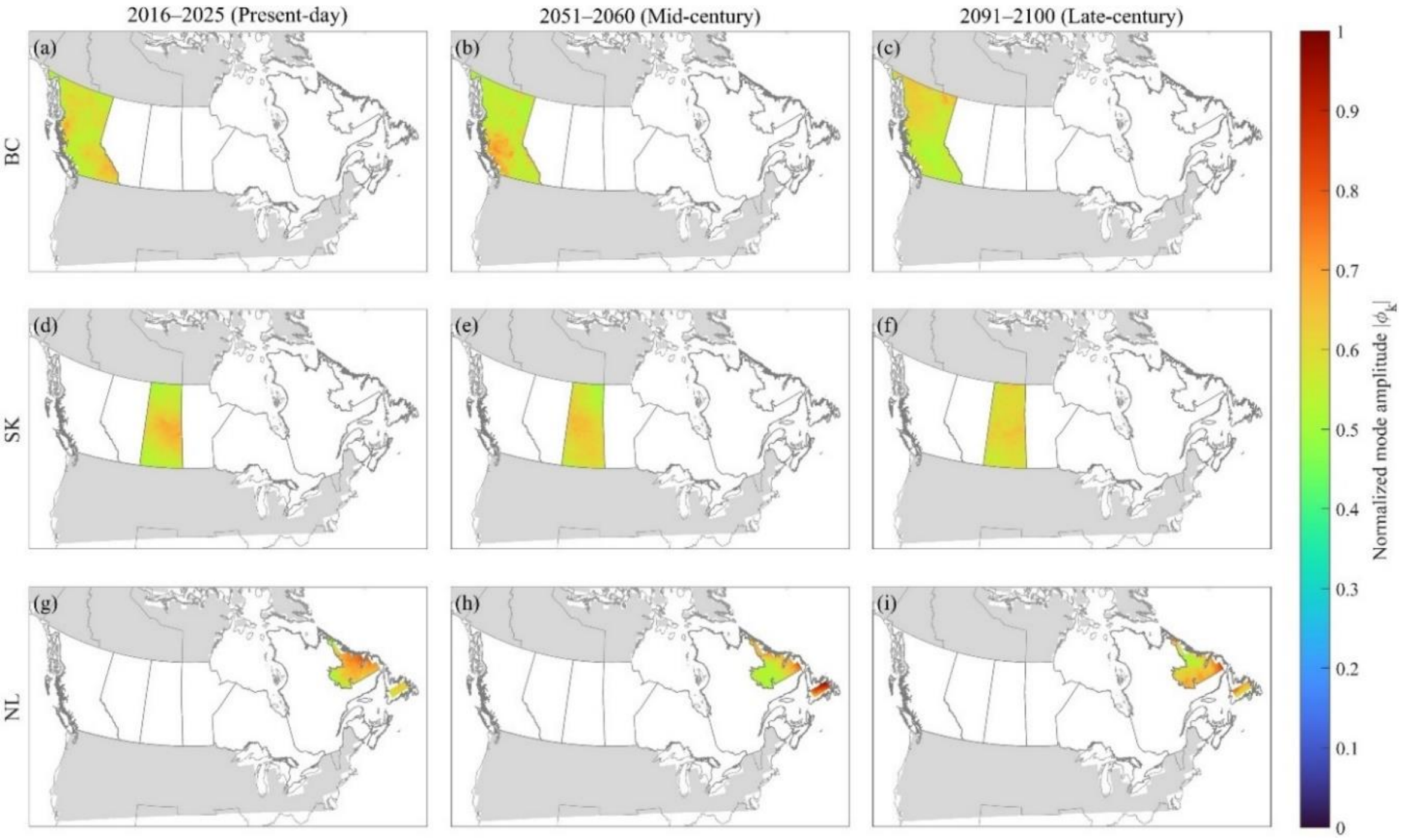

**Fig. 9** Spatial amplitude of the dominant low-frequency mrDMD mode across periods. Maps show normalized $|\phi_k|$ for the leading low-frequency mode in BC, SK, and NL during 2016–2025, 2051–2060, and 2091–2100.

### 3.4 Integrated attribution: linking mrDMD and heatwaves

This subsection links the event-based heatwave diagnostics (Section 3.2) with the multiscale dynamical structures extracted by mrDMD (Section 3.3) to identify which coherent modes are most relevant to heatwave behavior and to assess whether these modes organize the spatial geography of heatwave persistence and intensity. The attribution is implemented through two complementary diagnostics. First, a heatwave-conditioned participation ratio $P_k$ (Section 2.4; Eq. 8) ranks modes by preferential activation during heatwave days relative to non-heatwave conditions, thereby isolating structures that are dynamically engaged during exceedance episodes rather than simply energetic in an unconditional sense. Second, a spatial alignment analysis compares the energy footprints of the heatwave-relevant low-frequency modes against gridded heatwave metrics (HWMD and HWI in this study) using hotspot overlap and rank-based association (Spearman correlation), quantifying the extent to which the identified coherent structures explain the spatial organization of persistence and intensity hotspots.

### 3.4.1 Heatwave-conditioned mode participation

To identify dynamical structures that are preferentially engaged during heatwave conditions, the mrDMD modal coefficients are conditioned on heatwave days defined by the event-detection procedure in Section 2.2, using the integrated attribution framework in Section 2.4. For each retained mode $k$, heatwave relevance is quantified by the participation ratio $P_k$ (Eq. 8), defined as the ratio between the mean modal amplitude during heatwave days and the mean modal amplitude during non-heatwave days, where $A_k(t) = |b_k(t)|$ is the phase-independent modal amplitude. Modes with $P_k \gg 1$ therefore represent coherent structures whose activity increases disproportionately during exceedance conditions and are interpreted here as heatwave-relevant, in contrast to modes that may be energetic in an unconditional sense but exhibit comparable activity across heatwave and non-heatwave regimes.

For each region and period, modes are ranked by $P_k$ within the retained low-frequency band selected in Section 3.3.1, and a small subset of dominant heatwave-relevant modes is retained for interpretation. Throughout the main attribution analyses, interpretation is based on the top 5 heatwave-relevant modes. For the event-scale analysis in Fig. 10, however, event-integrated modal activity is computed from the top 10 modes to provide a broader cumulative representation of low-frequency activity. This subset captures a substantial fraction of the heatwave-day dynamical signal (explaining nearly 50% of the cumulative heatwave-day modal activity), indicating that heatwave conditions are associated with a relatively compact set of coherent low-frequency structures rather than a diffuse contribution across many modes.

Across periods, the participation spectrum evolves in a manner consistent with the persistence-driven regime shift reported in Section 3.2. In interior regions, the highest-participation modes are increasingly concentrated in the longest retained time scales, and the contrast between heatwave and non-heatwave activation strengthens toward the late century. Coastal and maritime-influenced regions exhibit comparatively weaker participation contrasts and a more distributed set of heatwave-relevant modes, consistent with their smaller duration amplification.

Figure 10 provides event-scale evidence linking heatwave persistence to heatwave-relevant dynamical activity in a representative interior region (SK). Figures 10a–c show the seasonal evolution of the amplitude of a selected low-frequency heatwave-relevant mode, with heatwave episodes shaded for each period. In the present-day climate, elevated modal amplitudes occur intermittently and are associated with relatively short heatwave episodes. By mid-century and especially by late century, the modal signal exhibits more frequent and stronger episodic amplifications during heatwave periods, with some bursts coinciding with longer events. In Fig. 10(c), the extended pink shading corresponds to a long continuous sequence of regional heatwave days in the selected late-century season used for illustration. This linkage is quantified in Fig. 10d, which relates event duration to event-integrated modal activity (summed over the top 10 heatwave-relevant modes). The positive but moderate association (Spearman $\rho = 0.45$) indicates that longer heatwaves tend to occur during periods of greater cumulative low-frequency modal activity.

These results suggest that duration amplification in interior regions is associated with recurrent activation of a small number of coherent low-frequency modes, while also indicating that additional processes beyond the modal activity shown here contribute to event persistence. This motivates the spatial attribution analysis in Section 3.4.2, which assesses whether the dominant heatwave-relevant modes also align with the geography of heatwave persistence and intensity hotspots.

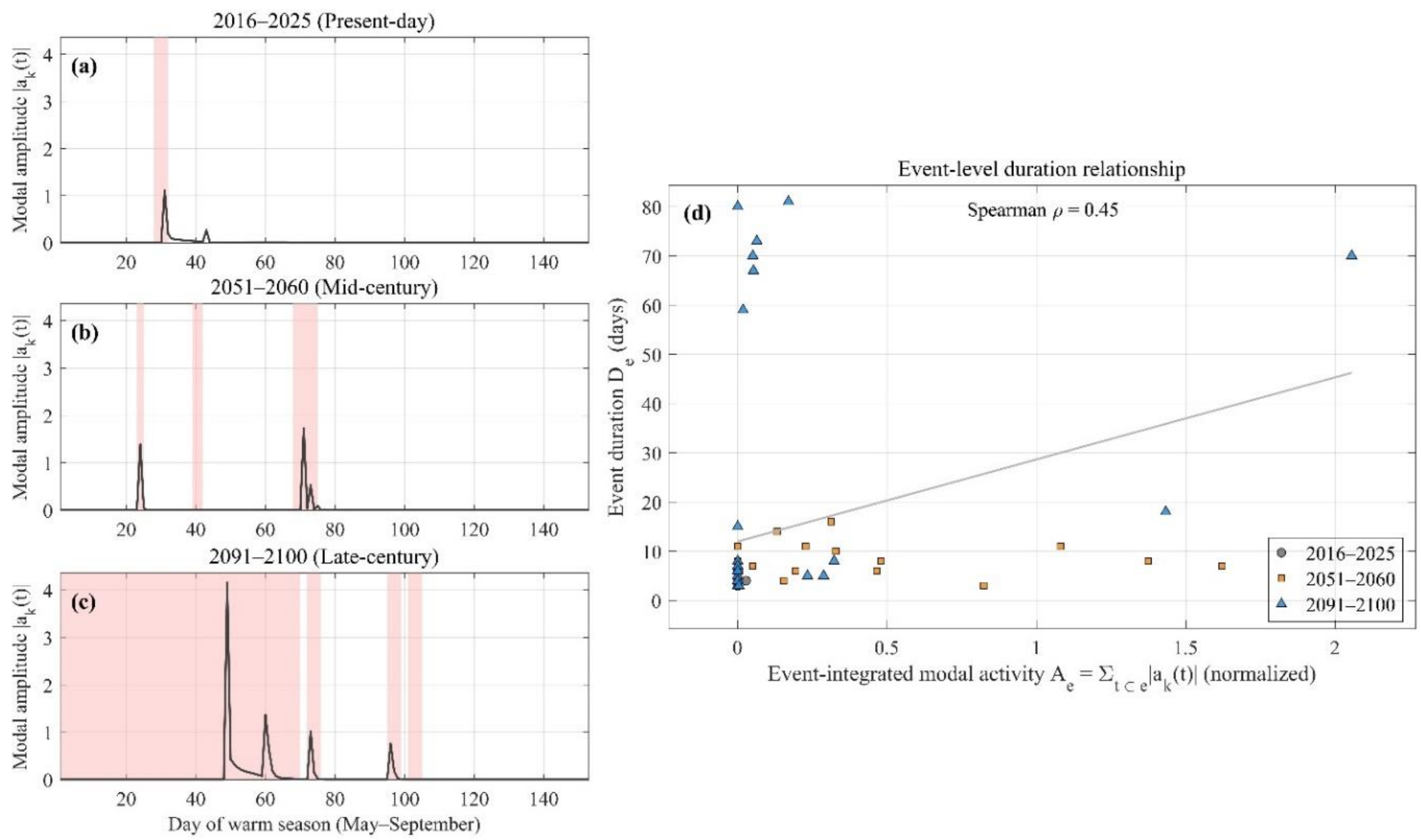

**Fig. 10** Event-scale linkage between heatwaves and heatwave-relevant dynamical modes. Seasonal evolution of the amplitude of the highest-participation low-frequency mrDMD mode in SK for (a) 2016–2025, (b) 2051–2060, and (c) 2091–2100, with heatwave episodes shaded; (d)

event duration versus event-integrated modal activity, showing that longer heatwaves are associated with more sustained modal activation.

### 3.4.2 Spatial alignment between heatwave metrics and heatwave-relevant mode footprints

While Section 3.4.1 identifies modes that are preferentially activated during heatwave days, the key remaining question is whether those same modes also organize the spatial geography of heatwave persistence and intensity. This is assessed by comparing gridded heatwave-metric maps with the spatial energy footprints of the heatwave-relevant low-frequency modes, as defined in Section 2.4. For each region and period, the analysis is based on the top 5 heatwave-relevant low-frequency modes identified by the participation analysis. This construction emphasizes spatial patterns expressed by modes that are not only low-frequency, but also demonstrably more active during exceedance conditions.

Figure 11 compares the resulting dynamical footprints against the spatial distribution of heatwave persistence and intensity. Panels 11a and 11e show gridded maps of persistence (HWMD) and intensity (HWI), respectively, while panels 11b and 11f show the corresponding composite footprint $E_{\mathrm{HW}}(x)$. The maps reveal a clear correspondence in the interior: regions exhibiting large HWMD and elevated HWI also tend to coincide with high-amplitude, spatially coherent dynamical footprints, consistent with the interpretation that persistent exceedance conditions preferentially emerge where the dominant heatwave-relevant low-frequency structures express most strongly. This correspondence is quantified in two complementary ways. First, a hotspot overlap analysis (Fig. 11c, g) evaluates whether upper-tail regions of the heatwave metrics coincide with upper-tail regions of the dynamical footprint. Hotspots are defined as grid cells exceeding a chosen upper percentile threshold (e.g., top $q = 20\%$) for each field, and overlap is summarized by an overlap score, where 0 indicates no overlap, and 1 indicates complete overlap. Second, Fig. 11(d, h) quantifies grid-cell association using Spearman rank correlation between $E_{\mathrm{HW}}(x)$ and each heatwave metric field. For the representative entire-domain case shown in Fig. 11, the spatial association is stronger for intensity (HWI) than for persistence (HWMD), indicating that the heatwave-relevant low-frequency footprint aligns more directly with the geography of intensity amplification in this case (HWMD, $\rho = 0.15$; HWI, $\rho = 0.45$). Coastal and maritime-influenced regions show weaker hotspot overlap and smaller rank correlations, consistent with their reduced persistence amplification and the smaller dynamical shifts identified in Sections 3.3.1–3.3.3.

Overall, Fig. 11 provides spatial evidence that the modes identified as heatwave-relevant by participation encode the spatial organization of heatwave impacts. Notably, the spatial alignment is moderate for event intensity ($\rho = 0.45$), while the association with event persistence is more limited ($\rho$ = 0.15), suggesting that while these modes temporally sustain heatwaves, their spatial energy footprints most directly govern the magnitude of the extreme temperatures.

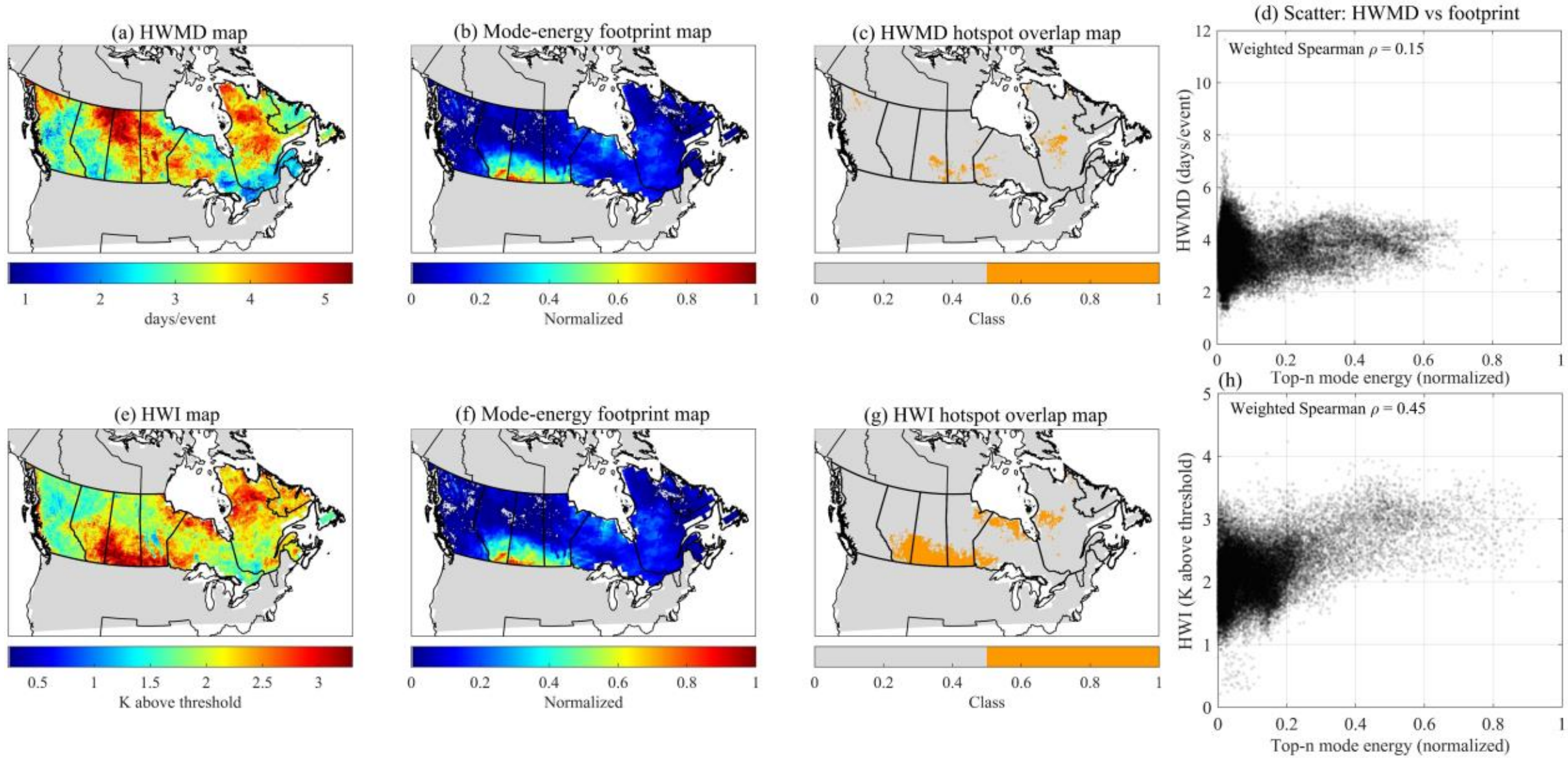

**Fig. 11** Spatial alignment between heatwave metrics and heatwave-relevant dynamical mode footprints for the representative late-century (entire Canada domain). Maps compare (a,e) heatwave persistence (HWMD) and intensity (HWI) with (b,f) the composite footprint of heatwave-relevant low-frequency mrDMD modes; (c,g) show hotspot overlap (upper-tail regions), and (d,h) quantify grid-cell association using Spearman rank correlation.

Taken together, the participation and spatial-alignment results indicate that late-century duration amplification is not merely a statistical consequence of warming, but is associated with a systematic strengthening of specific low-frequency dynamical structures. In interior regions, these modes (i) exhibit weaker damping and increased low-frequency dominance (Section 3.3), (ii) activate preferentially during heatwave days and remain elevated through multi-day events (Fig. 10), and (iii) align spatially with persistence and intensity hotspots (Fig. 11). Coastal and maritime regions show weaker dynamical shifts and weaker footprint–metric alignment, consistent with their smaller persistence amplification. These findings support a dynamical interpretation of the projected heatwave regime shift in which persistent exceedance emerges from increasingly coherent, weakly damped low-frequency organization of warm-season temperature variability.

## 4. Discussion

### 4.1 Interpretation of the projected heatwave regime shift

The results indicate that projected heatwave amplification under SSP5–8.5 is not adequately described as a uniform increase in exceedance frequency alone. Instead, the most consequential change, particularly over the continental interior, takes the form of a regime shift toward persistence-dominated exceedance. Event-based diagnostics show that future increases in seasonal heatwave days arise from a combined effect of more frequent exceedances and, critically, longer event durations, with the largest persistence amplification occurring over interior regions (Fig. 5). This shift is reflected in the dynamical decomposition: the mrDMD representation reveals a progressive redistribution of dominant activity toward lower-frequency levels in the interior relative to coastal and maritime regions (Fig. 7), accompanied by eigenvalue distributions indicative of weaker damping and longer e-folding time scales (Fig. 8).

The integrated attribution analysis provides a consistent mechanistic interpretation of these statistical changes. Modes that are preferentially activated during heatwave days are concentrated in the retained low-frequency band and exhibit increasingly sustained amplitudes through multi-day events in late-century conditions (Fig. 10). Moreover, the spatial energy footprints of these heatwave-relevant modes align with the geography of persistence and intensity hotspots (Fig. 11), indicating that the coherent structures identified by mrDMD are not merely correlated with heatwave timing but also encode the spatial organization of heatwave impacts. Taken together, the evidence supports the interpretation that late-century duration amplification is associated with an increasing dominance of slowly varying, weakly damped structures that promote sustained exceedance once heatwave conditions are initiated.

This perspective helps reconcile why similar levels of mean warming can yield markedly different regional heatwave outcomes. Regions where the low-frequency dynamical background becomes more coherent and weakly damped are more prone to prolonged exceedance sequences, even if changes in event counts are comparatively modest. Conversely, coastal and maritime-influenced regions exhibit weaker shifts in low-frequency dominance and damping, consistent with smaller persistence amplification and weaker footprint–metric alignment. These contrasts suggest that regional differences in heatwave risk under climate change reflect not only thermodynamic forcing (warmer baseline temperatures), but also changes in the dynamical organization of warm-season variability that modulate persistence.

From an impact's perspective, the dominance of persistence changes has direct relevance for risk assessment and adaptation planning. Longer events compound health and infrastructure stress through cumulative exposure, reduce nighttime recovery opportunities, and increase the likelihood of concurrent stresses (e.g., drought and wildfire risk) during extended warm-season anomalies. The dynamical framing provided here therefore complements conventional event-statistical projections by identifying the time-scale structures most responsible for sustained exceedance, offering a pathway to more process-informed diagnostics of future heat risk and to targeted evaluation of whether climate models reproduce the relevant low-frequency persistence mechanisms in the historical climate.

**4.2 Implications, robustness, and limitations**

The integrated event–dynamical framework provides a compact way to move from projected heatwave changes to a physically interpretable explanation rooted in the time-scale organization of warm-season temperature variability. A key implication is that persistence is an emergent property of the dominant low-frequency background, not simply a by-product of higher mean temperatures. In the interior, the attribution results indicate that duration amplification is associated with increasingly weakly damped, spatially coherent low-frequency structures that are preferentially activated during heatwave days and that align with regional persistence and intensity hotspots (Figs. 8–11). This suggests that assessments of future heat risk, and especially adaptation-relevant questions such as multi-day exposure and compound impacts, should place as much emphasis on diagnosing changes in low-frequency variability and persistence as on changes in exceedance frequency alone.

The framework also offers practical diagnostic targets for model evaluation. Conventional validation of temperature statistics (means, variances, percentile thresholds) is necessary but may be insufficient for persistence-dominated hazards. The present results point to additional "process-level" checks, including whether historical simulations reproduce (i) the distribution of damping time scales $\tau_k$ for warm-season low-frequency modes, (ii) the relative contribution of low-frequency levels to warm-season variability (as in Fig. 7), and (iii) the event-scale linkage between sustained modal activation and long-duration heatwaves (as in Fig. 10). These diagnostics can be applied across climate model ensembles to determine whether persistence amplification is robust across model formulations and to identify whether inter-model spread is tied to differences in low-frequency variability rather than to differences in mean warming alone.

Several limitations and robustness considerations should be noted. First, the analysis is based on a single downscaled model chain (CRCM5 driven by CanESM5) and a single forcing pathway (SSP5–8.5). While SSP5–8.5 provides a high-signal scenario for diagnosing mechanisms, the quantitative magnitude of duration amplification and the strength of dynamical shifts may differ under lower-emissions pathways. Second, the event definition is tied to a fixed historical percentile threshold; this choice is appropriate for identifying changes relative to a consistent present-day, but alternative definitions (e.g., thresholds based on future climatologies, absolute health-relevant thresholds, or humidity-adjusted indices) may yield different perspectives on risk. Third, some methodological choices influence interpretability and should be reported transparently. These include the retained mrDMD cut-off parameter $F_c$ (Fig. 6), the truncation rank $r$, and the criteria used to identify heatwave-relevant modes (top- $K$ selection). The sensitivity analysis in Section 3.3.1 suggests that the principal conclusions are not artifacts of a single cut-off choice, but the numerical values of correlations and hotspot overlap may vary with these settings. Reporting the selected parameters and providing limited sensitivity results (e.g., alternative $F_c$ values) strengthens reproducibility and makes clear which aspects of the attribution are robust versus configuration-dependent.

Overall, the results support a shift in emphasis from purely event-statistical projections toward combined diagnostics that explicitly track the dynamical structures governing persistence. This provides a pathway to (i) more process-informed interpretation of regional heatwave amplification, (ii) improved comparability across models and scenarios, and (iii) targeted evaluation of whether the variability structures that matter most for long-duration heat risk are credibly represented in climate simulations.

**5. Conclusion**

This study developed an integrated event–dynamical workflow to assess how climate change alters regional heatwave risk and to interpret projected amplification in terms of the multiscale organization of warm-season temperature variability. Heatwaves were characterized using percentile-based event diagnostics that quantify frequency, duration, and intensity, and multiresolution dynamic mode decomposition (mrDMD) was then used to extract coherent spatio-temporal structures from the raw daily mean temperature field. Linking these perspectives through heatwave-conditioned modal activation and spatial alignment diagnostics enables a transition from documenting projected changes to identifying the dynamical structures most strongly associated with those changes. The results indicate that projected heatwave amplification under SSP5–8.5 is characterized not only by more frequent exceedance, but by a pronounced intensification of persistence in the continental interior. In these regions, increases in seasonal heatwave days are accompanied by substantially longer events, pointing to a shift toward persistence-dominated exceedance regimes. Dynamical diagnostics provide consistent evidence for this shift: interior regions exhibit a redistribution of dominant activity toward lower-frequency content together with weaker effective damping, while coastal and maritime regions show comparatively smaller changes in low-frequency dominance and persistence. Integrated attribution further shows that a compact set of low-frequency modes becomes preferentially activated during heatwave days, remains elevated through long events, and exhibits spatial energy footprints that align with the geography of persistence and intensity hotspots. Taken together, these findings support a dynamical interpretation in which late-century duration amplification reflects increasing dominance and spatial coherence of weakly damped low-frequency structures that sustain prolonged exceedance once heatwave conditions are initiated. Beyond the specific case study, the framework provides process-informed diagnostics that complement conventional event-statistical projections. By explicitly isolating the time-scale structures governing persistence, it offers actionable targets for model evaluation and a basis for comparing persistence mechanisms across regions, scenarios, and model ensembles. Future extensions include multi-model and multi-scenario applications to quantify robustness and uncertainty, and conditioning on impact-relevant heat metrics to connect persistence-oriented dynamics more directly to health and infrastructure risk.

## Appendix

Figures S1–S2 provide supplementary summaries that support the regional heatwave projections and the integrated dynamical attribution results presented in the main text. All metrics are computed over the May–September season using the fixed historical threshold and event definition described in Section 3.1. Fig. S1 reports the regional distributions of heatwave metrics across the eight subregions (BC, AB, SK, MB, ON, QC, NL, NB) for 2016–2025, 2051–2060, and 2091–2100. For each metric, values are first computed at the grid-cell level and then aggregated to each subregion using area-weighted averaging. Distributions summarize interannual variability within each 10-year period (one regional value per year).

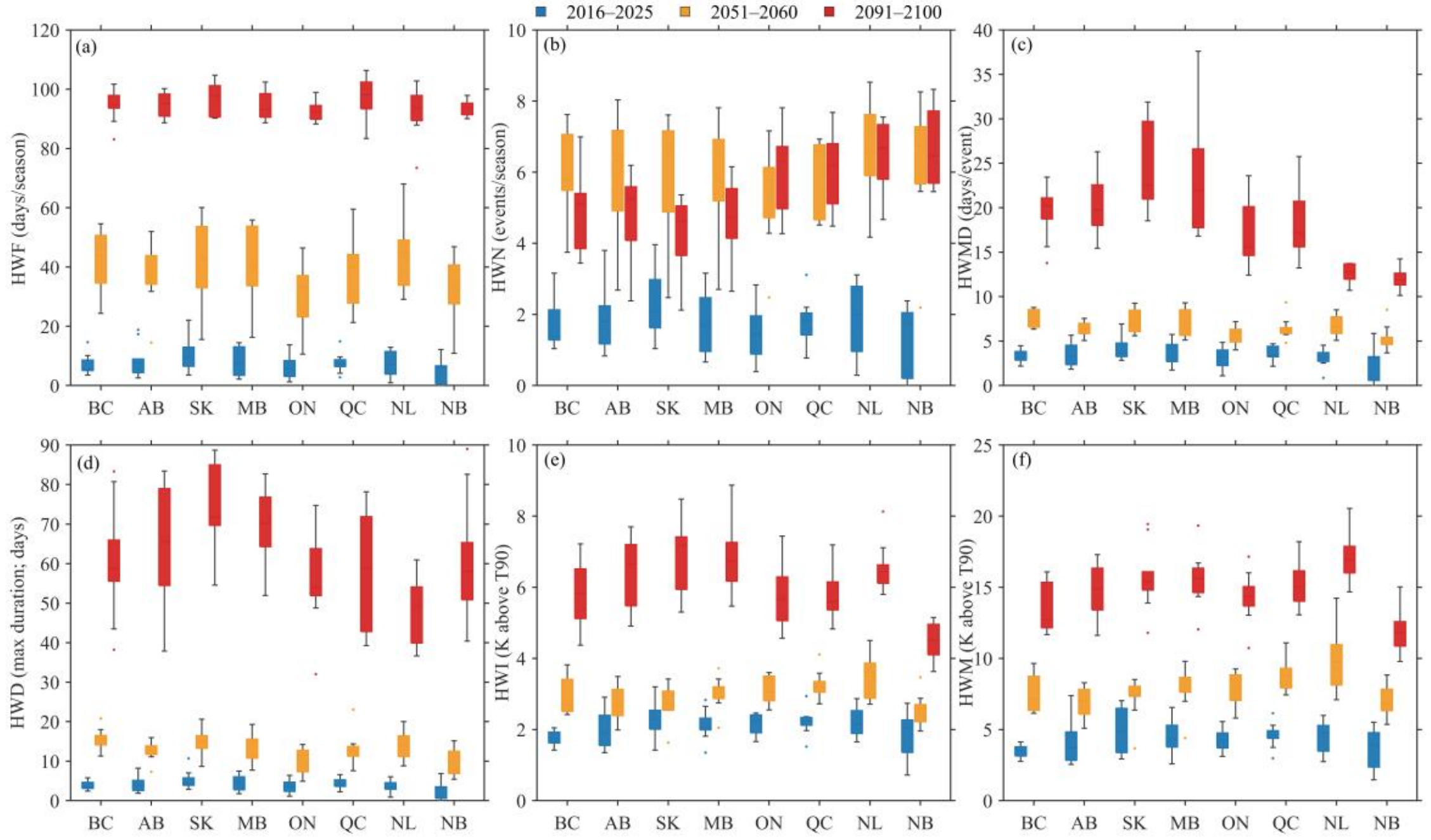

**Fig. S1** Regional distributions of heatwave metrics (May–September). Distributions are shown for BC, AB, SK, MB, ON, QC, NB, and NL during 2016–2025, 2051–2060, and 2091–2100: (a) HWF (days/season), (b) HWN (events/season), (c) HWMD (days/event), (d) HWD (maximum event duration; days), (e) HWI (mean exceedance intensity; K above threshold), and (f) HWM (maximum exceedance intensity; K above threshold).

Fig. S2 provides a compact cross-region summary of the spatial alignment between gridded heatwave metrics and the composite energy footprint of heatwave-relevant low-frequency modes. For each subregion and period, the footprint constructed from the top-$K$ heatwave-relevant modes ($K = 5$) is compared against each metric map using (i) an area-weighted Spearman rank correlation and (ii) a hotspot overlap score based on the intersection of upper-tail grid cells. This summary is included to facilitate comparison of alignment strength across metrics, regions, and periods.

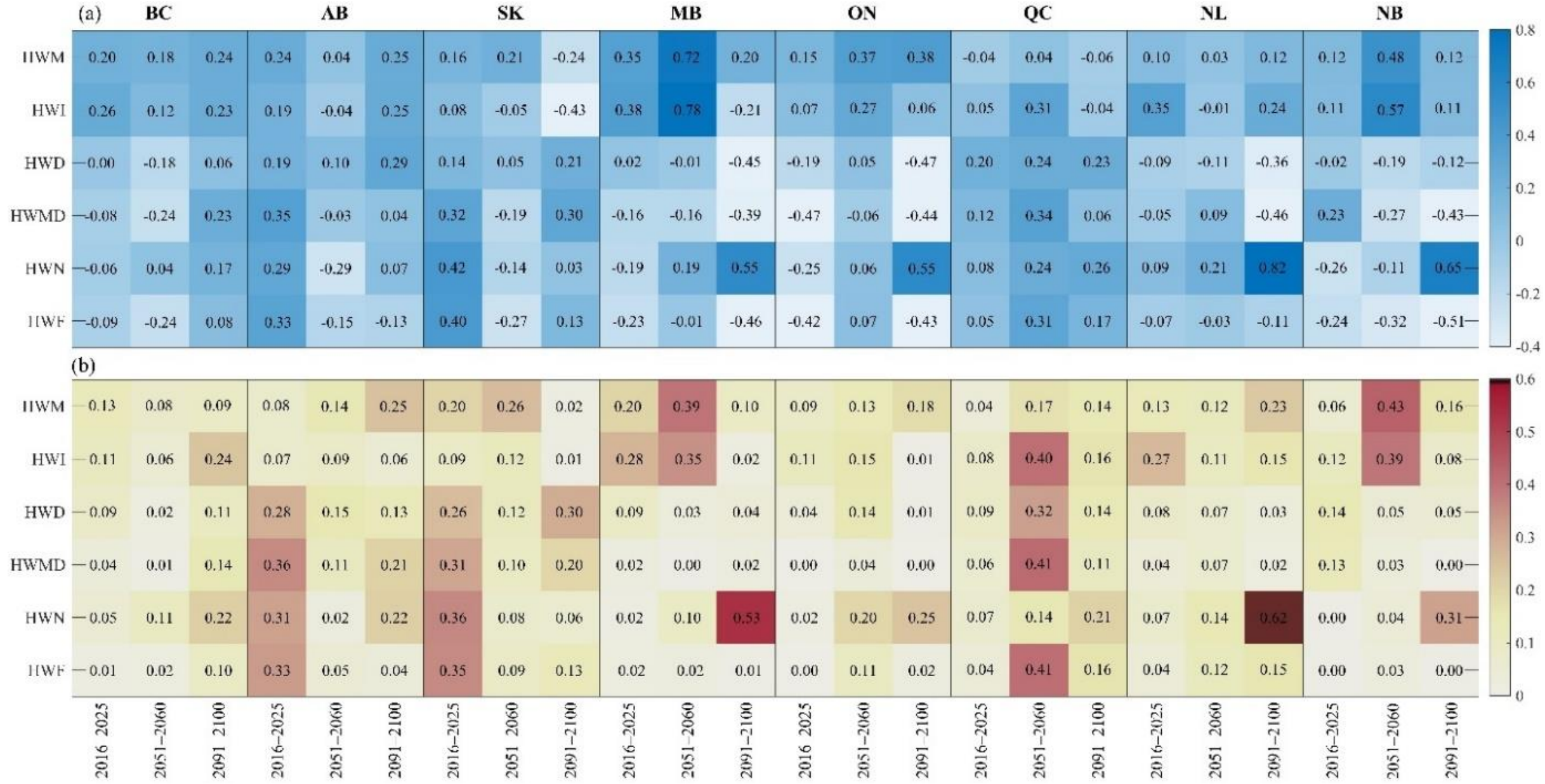

**Fig. S2** Province-scale summary of spatial alignment between heatwave metrics and heatwave-relevant mode footprints. (a) Area-weighted Spearman rank correlation $\rho$ between gridded heatwave-metric maps and the composite mode-energy footprint derived from the top-$K$ heatwave-relevant low-frequency modes, reported for each subregion and period. (b) Corresponding hotspot overlap score based on the intersection of upper-tail regions (top $p$=20%).

## Declaration of Competing Interest

The authors declare that they have no known competing financial interests or personal relationships that could have appeared to influence the work reported in this paper.

## Acknowledgements

This project was supported by École de Technologie Supérieure (ETS) Startup Fund (number 102690) and the Natural Sciences and Engineering Research Council of Canada (NSERC) (grant number RGPIN 2022–03492)